\documentclass[computers,article,accept,moreauthors]{Definitions/mdpi} 

\usepackage{amsmath,amssymb,amsfonts}
\usepackage{algorithmic}
\usepackage{multicol}
\usepackage{graphicx}
\usepackage{caption}
\usepackage{textcomp}
\usepackage{multirow} 
\usepackage{subfig}
\usepackage{hyperref}

\def\BibTeX{{\rm B\kern-.05em{\sc i\kern-.025em b}\kern-.08em
    T\kern-.1667em\lower.7ex\hbox{E}\kern-.125emX}}

\usepackage{xcolor} 
\definecolor{blue}{RGB}{51,153,255}

\firstpage{1} 
\pubvolume{9}
\issuenum{3}
\articlenumber{64}

\pubyear{2020}
\copyrightyear{2020}
\history{Received: 14 July 2020; Accepted: 9 August 2020; Published: 12 August 2020}

\Title{Privacy Preserving Passive DNS}

\Author{Pavlos Papadopoulos $^{1,}$*, Nikolaos Pitropakis $^{1}$*, William J. Buchanan $^{1}$, Owen Lo $^{1}$ and Sokratis Katsikas $^{2,3}$}

\AuthorNames{Pavlos Papadopoulos, Nikolakos Pitropakis, William J. Buchanan, Owen Lo and Sokratis Katsikas}

\address{%
$^{1}$ \quad School of Computing Edinburgh Napier University, Edinburgh, United Kingdom; pavlos.papadopoulos@napier.ac.uk (P.P.); N.Pitropakis@napier.ac.uk (N.P.); B.Buchanan@napier.ac.uk (W.J.B.); O.Lo@napier.ac.uk (O.L.)\\
$^{2}$ \quad Department of Information Security and Communication Technology, Norwegian University of Science and Technology, Norway; sokratis.katsikas@ntnu.no (S.K.)\\
$^{3}$ \quad Faculty of Pure and Applied Sciences, Open University of Cyprus, Cyprus; sokratis.katsikas@ouc.ac.cy (S.K.)}

\corres{Correspondence: \{N.Pitropakis,pavlos.papadopoulos\}@napier.ac.uk}

\abstract{The Domain Name System (DNS) was created to resolve the IP addresses of the web servers to easily remembered names. When it was initially created, security was not a major concern; nowadays, this lack of inherent security and trust has exposed the global DNS infrastructure to malicious actors. The passive DNS data collection process creates a database containing various DNS data elements, some of which are personal and need to be protected to preserve the privacy of the end users. To this end, we propose the use of distributed ledger technology. We use Hyperledger Fabric to create a permissioned blockchain, which only authorized entities can access. The proposed solution supports queries for storing and retrieving data from the blockchain ledger, allowing the use of the passive DNS database for further analysis, e.g. for the identification of malicious domain names. Additionally, it effectively protects the DNS personal data from unauthorized entities, including the administrators that can act as potential malicious insiders, and allows only the data owners to perform queries over these data. We evaluated our proposed solution by creating a proof-of-concept experimental setup that passively collects DNS data from a network and then uses the distributed ledger technology to store the data in an immutable ledger, thus providing a full historical overview of all the records. }


\begin{document}

\section{Introduction}
\label{sec:introduction}

\begin{figure}[h!]
\centering
\includegraphics[width=0.8\linewidth]{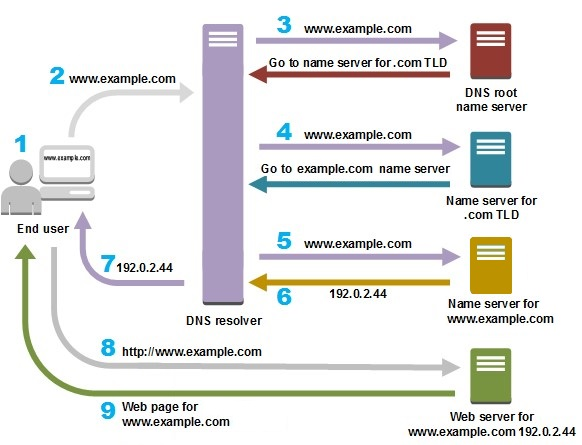}
\caption{Overview of the DNS}
\label{DNSoperation}
\end{figure}

The Domain Name System (DNS) translates human-readable domain names to machine-readable IP addresses \cite{mockapetris1987domain}. An end-user can thus access a website through a web browser using a combination of a name e.g. ``example'' and a TLD e.g. ``.com'', ``.uk'', ``.us''. Figure \ref{DNSoperation} illustrates the steps involved in the DNS query resolution process. DNS provides a foundation element of the trustworthiness of the Internet, but its simplicity and general lack of trust has led to a range of security issues. Botnets \cite{antonakakis2017understanding}, parking domains \cite{vissers2015parking} and domain squatting \cite{stout2012system} are examples of types of malicious DNS use. To identify DNS abuse, the DNS queries and responses often have to be collected for further analysis.

Florian Weimer is the creator of the Passive DNS systems \cite{weimer2005passive}; he used recursive name servers to log responses received from different name servers and then copied this logged data onto a central database. The almost instantaneous recording of the majority of passive DNS data before the recursive name server means that Passive DNS is composed of responses and referrals from online authoritative name servers. These logged data are deduped, compressed, time-stamped, and then copied onto a central database where they are analysed and archived. 

Passive DNS collects the DNS queries along with the IP address of the host that is making the queries. In situations when the passive DNS collector is placed within the ISP (Internet Service Provider) or at a TLD (Top-Level Domain) server, each query contains the IPs of the end-users and can be linked back to them. Both the GDPR and NIST consider IP addresses as personal data when a correlation of the queries and the identity of the end-user can be made \cite{GDPRip,spring2012impact}. When it comes to public DNS servers, end-users can benefit from better (than the DNS servers of their ISP) stability, availability and protection against certain DNS attacks, but they expose their personal data to companies such as Google \cite{Googledns}, Cloudflare \cite{Cloudflaredns}, and OpenDNS \cite{Opendns} that could profit from commercially exploiting these data \cite{federrath2011privacy}. \textcolor{black}{It follows that there is a clear need to appropriately protect data collected, stored, and processed to identify malicious domain names through the use of passive DNS data analysis methods. The majority of existing solutions for passive DNS data analysis provide APIs for queries of the related data. However, the collection of passive DNS data is being questioned \cite{khalil2016discovering}, since the privacy of the end-users that contributed their passive DNS collections may be compromised \cite{gasser2018log}.} 

This paper proposes a solution to the aforementioned problem and outlines \textbf{P}rivacy P\textbf{reser}ving Passi\textbf{ve} (\textit{\textbf{PRESERVE DNS}}), a system that collects passive DNS data for further analysis, whilst preserving the privacy of the end-users, by virtue of storing the data in an immutable distributed ledger. This immutable ledger can be used to identify abuses and malicious DNS usage, such as for domain squatting and botnets. Blockchain technology can also provide the complete history of the data transactions, without the need for a central authority to control it. The proposed solution is not connected with any passive DNS database provider, such as Farsight (DNSDB) \cite{Farsightdnsdb}, and VirusTotal \cite{Virustotalpassivedns}). Furthermore, it provides transparency of the stored data, thus establishing trust with the users, while at the same time allowing authorized entities to query only non-personal data.

Not all data collected are personal or in need of the same level of privacy protection. Selectively protecting personal data, such as source IP addresses, is possible in a permissioned distributed ledger. In a permissioned-protected field of an immutable ledger, even the administrators do not have access to the private data. This choice allows only the data owners to query the data. In this paper, Hyperledger Fabric is the selected permissioned blockchain platform. This platform has the advantage of resolving potential scalability issues since consensus of participating peers and their respective permissions can be configured. Additionally, with the encryption and anonymisation functionality provided within Hyperledger Fabric, participating entities can securely provide passive DNS collections to the system.  

The contributions of our work can be summarised as follows:
\begin{itemize}
\item  We have developed PRESERVE DNS, a privacy preserving passive DNS data solution, by leveraging distributed ledger technology. The implementation of this solution does not require any changes to the server side of the current DNS infrastructure.
\item We have evaluated the robustness and security of PRESERVE DNS.
\item We have comparatively evaluated the performance of PRESERVE DNS against a traditional database with column level encryption, and against an existing alternative solution.
\end{itemize}

The remaining of the paper is organised as follows: Section~\ref{Background} provides the background on blockchain technology that is necessary to ensure the self-sustainability of the paper. The relevant literature is discussed in Section~\ref{Relatedwork}. In Section~\ref{PrivacyPreservingPassiveDNS} the proof-of-concept implementation of PRESERVE DNS, which also serves as the evaluation testbed, are discussed. Section~\ref{evaluation} presents the process and the results of the evaluation of PRESERVE DNS. Finally, Section~\ref{ConclusionsandFutureWork} summarizes our conclusions and suggests some future work.

\section{Background} 
\label{Background}
\subsection{\textcolor{black}{DNS Privacy Concerns}}
\textcolor{black}{The DNS infrastructure is outdated and has been created without considering security or privacy. This leads to DNS being targeted by numerous malicious actors that try to exploit its vulnerabilities to get profit from unaware end-users. Malevolent individuals are able to profit directly from their victims, or by selling batches of their private information \cite{james2005phishing,yoon2019doppelgangers,riederer2011sale}. A number of current privacy issues cannot be fully resolved without a complete redesign of DNS \cite{bortzmeyer2015dns}. Fortunately, new solutions are being proposed and developed to further enhance the security of DNS and to preserve the privacy of its end-users. Blockchain DNS solutions promise to resolve existing DNS privacy issues \cite{wang2019survey,benisi2020blockchain,liu2018data}. Our work combines part of the old DNS infrastructure with the new blockchain technology.}

\subsection{Blockchain}
\label{blockchain}
Blockchain is a distributed ledger technology \cite{di2017blockchain}, that became
popular as the foundational block of the \textit{bitcoin} cryptocurrency \cite{nakamoto2019bitcoin,zheng2017overview}. Over the past few years it has seen rapid growth, both in terms of research and commercial usage. \textcolor{black}{A \textit{blockchain} is a cryptographically-linked list of records that maintains a publicly verifiable ledger without the need for a central authority; as such, it is a new paradigm of trust between entities in various application domains. The technology behind blockchains originated in cryptocurrency applications but, its advancements over existing architectures motivated researchers to apply it to a broad spectrum of application domains \cite{dai2019blockchain,luo2020blockchain,xu2019apis,wu2019comprehensive,karandikar2019transactive}}. The main benefits of this technology are the decentralized nature, immutability, inherent anonymity, resilience, trust, security, autonomy, integrity and scalability.

Blockchains can be categorized according to who can access them as \textit{public}, \textit{private} or \textit{consortium}. Anyone can join a \textit{public} blockchain and act as a simple node or as miner/validator; no approval by any third party is needed. In the case of a \textit{private} \textcolor{black}{or \textit{consortium}} blockchains, the owner restricts network access, \textcolor{black}{to herself or to a group of participants}. Blockchains can also be categorized according to whether each entity requires authorization to perform an action or not as \textit{permissioned} or \textit{permissionless} respectively \cite{xu2017taxonomy}.    

\subsection{Hyperledger Fabric}
\label{hlftopologysection}
Hyperledger Fabric is an open source enterprise-grade permissioned \textcolor{black}{distributed ledger technology} platform \cite{androulaki2018hyperledger}. The main participants in a Hyperledger Fabric blockchain are the \textit{peers}, the \textit{organizations}, the \textit{orderers}, the \textit{Certificate Authority}, and the \textit{Membership Service Provider (MSP)}. Participants perform actions on the ledger by using \textit{chaincode} \cite{karandikar2019transactive}.

The \textit{chaincode} is a blockchain program that runs autonomously and performs a set of actions defined by the developer. It is written in common general-purpose programming languages such as Javascript, Java or Go. The chaincode is installed and instantiated in the peers and the orderer, and contains all the blockchain's logic, security mechanisms and capabilities. 

\textit{Peers} are the most crucial entities on the blockchain network. They install the chaincode, and they host the ledger. Peers can host more than one ledgers and chaincodes, enabling private communications with other entities. 

\textit{Organizations} are groups of peers whose actions can be defined by policy \cite{Hyperledger}. The network is formed by different organizations, that all together form a consortium \cite{dhillon2017blockchain}. It is possible that different ledgers can be present at the same time and only authorized organisations and entities have access to them.

\textit{Ordering Service} or \textit{orderer} is the entity that receives the transactions from the peers' applications and updates the ledger according to the defined consensus mechanism. The \textit{consensus} mechanism can be defined during the time of creation, and complex fault-tolerant algorithms can be used for the validation of each transaction \cite{zhao2019consensus}. 

The \textit{Certificate Authority} is the entity that assigns an identity to each peer. The \textit{Membership Service Provider (MSP)} is the entity that validates the identity of each participant in the blockchain network. It manages and examines all the cryptographic mechanisms and certificates that the peers use to perform actions in the ledger \cite{androulaki2018hyperledger}.

\section{Related Work} 
\label{Relatedwork}

Despite the success of Weimer's concept, the issue of the impact of the collection of passive DNS data to end-user privacy was soon raised, as users could be clueless if a passive DNS collector is placed in their DNS resolver \cite{spring2012impact}. Consequently, the security research community focused on this issue and several approaches addressing the issue were published.  

One of the first approaches to mitigate the impact of the issue was to use tools that could eliminate confidential information from collected network packets \cite{zdrnja2006security}. Another approach argued that a Cryptography-based Prefix preserving Anonymization algorithm \cite{govil20074g} or other encryption techniques that would secure the IP prefix \cite{xu2002prefix} should be employed. At the other end of the spectrum, an entirely different solution was proposed: the collection of \textit{active} DNS data \cite{kountouras2016enabling}. This was made possible by creating a system called Thales which can systematically query and collect large volumes of active DNS data using as input an aggregation of publicly accessible sources of domain names and URLs that have been collected for several years by the research team. These sources include but are not limited to Public Blacklists, the Alexa ranking, the Common Crawl project, and various Top Level Domain (TLD) zone files. This system's output is a refined dataset that can be easily used by the security community. 

Liang et al. \cite{liang2017provchain} proposed a system that combines two technologies, namely blockchain and cloud computing to effectively and efficiently create a decentralised DNS records database. To ensure the security of the stored data, they employed a hashed version of the sensitive data as a proof-of-identity, and allowed only the administrator of the system to correlate each identity to the hashed data. This is a successful countermeasure against outsider malicious actors, but sensitive data may still be compromised when attacks are launched by insiders. In contrast, the data in PRESERVE DNS are stored by the users themselves and are available in a separate ledger only to them; the remaining network only have access to a hash of the actual data.

Liu et al. \cite{liu2018data} proposed a decentralized, blockchain-based DNS (DecDNS) system which maintains a stored database of DNS records and performs the resolution using the nodes of the blockchain. The advantages and default security mechanisms of blockchain, such as the tampered-proof state of the data and the resilience against Distributed Denial of Service (DDoS) attacks are important features of the system. Moreover, the solution does not require significant changes to the existing DNS infrastructure. However, the privacy of the users is not preserved; the sacalability of the solution is questionable; and performance is a challenge, as all the data are in a hashed form, and they should be decrypted in every DNS query. In contrast, in PRESERVE DNS public data needed for further analysis are in plain text, and only authorized entities are allowed to query them. 

Rather than attempting to secure the existing DNS infrastructure, another line of research proposes the development of a more secure, easily audited, transparent domain names organization. Examples are systems such as Namecoin \cite{kalodner2015empirical} and Blockstack \cite{ali2016blockstack}, that created a substitute of the Internet Corporation for Assigned Names and Numbers (ICANN), where each user does not need to buy a domain name from a third party. The proposed system is built on a blockchain network, bitcoin in this case, where users can ``mine'' cryptocurrency. Then users are able to use this cryptocurrency to buy domain names with new ``.bit'', ``.id'' TLDs that did not exist before \cite{ali2016blockstack}. The privacy of the users can be ensured since their identity is protected by bitcoin's identity management mechanism. The downside of these systems is that the users need specific extensions to be able to query blockchain registered domain names.

Along similar lines, Kalodner et al. \cite{kalodner2015empirical} and Ali et al. \cite{ali2016blockstack} proposed solutions that can address common DNS issues and attacks, by changing the existing infrastructure to a more secure, resilient version with far more opportunities, security mechanisms and defences. However attractive these proposals may seem, the requirement to make changes to the existing DNS global infrastructure is rather unrealistic. In contrast, PRESERVE DNS is able to secure the existing DNS functionality without the need for major changes at the server, thus offering the opportunity for a faster transition without downtime or enormous expenses. 

DNS Trusted Sharing Model (DNSTSM) is a recently published approach, which also uses the Hyperledger Fabric platform \cite{yu2020dnstsm}. DNSTSM is a system resilient to various DNS attacks, that achieves a high performance of DNS resolutions, and does not require changes in the current global DNS infrastructure. However, DNSTSM is built on the older v1.1 version of Hyperledger Fabric, which does not have the private data collection feature as the v1.4 version that PRESERVE DNS is built upon does. This means that DNSTSM cannot preserve end-user privacy without effectively re-designing its architecture so as to be able to exploit features available in newer versions of Hyperledger Fabric. 

PRESERVE DNS differentiates itself from previous works in various ways. By leveraging  the private data collection feature provided by Hyperledger Fabric, in PRESERVE DNS two separate ledgers are created, one for the public DNS information used for further analysis, and one for the sensitive data. The latter is stored only on the peer nodes of the owners of the data, and only they can query it. Furthermore, PRESERVE DNS stores the passive DNS data collection through storing API requests directly from the users, making use of each user's identity. This method enhances the privacy of the users as only themselves have access to their personal data and are able to query them. Additionally, PRESERVE DNS is efficient, since the rest of the data are in plain text, and only trusted validated users are allowed to query them. Data are available through query API requests to authorized entities, and further analysis of passive DNS data towards e.g. malicious domain name identification and domain squatting is possible. 

Finally, as will be discussed in section~\ref{evaluation}, PRESERVE DNS is able to thwart various DNS attacks such as DDoS, DNS fast-flux, DNS amplification attacks. In situations where the PRESERVE DNS distributed DNS records database is being used for the DNS resolution, the DNS cache poisoning attack, one of the most difficult attacks to defend against, can be thwarted as well.

A summary of salient characteristics of PRESERVE DNS and of the approaches discussed above is depicted in Table~\ref{tab:comparison}. 

\bgroup
\def\arraystretch{1.3}
\begin{table*}[h]
\caption{Comparison of Methods}
\centering
\begin{tabular}{|c | c| c| c|} \hline 
\textit{Method} & \textit{Attack Thwarting} & \textit{User privacy} & \textit{Existing DNS infrastructure} \\ [0.5ex] \hline
\textbf{DecDNS Liu et al. \cite{liu2006dns}} & \checkmark & \textbf{X} & \checkmark \\  
\textbf{Liang et al. \cite{liang2017provchain}} & \checkmark & \textbf{X} & \checkmark \\  
\textbf{Namecoin Kalodner et al. \cite{kalodner2015empirical}} & \checkmark & \checkmark & \textbf{X} \\  
\textbf{Blockstack Ali et al. \cite{ali2016blockstack}} & \checkmark & \checkmark & \textbf{X} \\
\textbf{DNSTSM Yu et al. \cite{yu2020dnstsm}} & \checkmark & \textbf{X} & \checkmark \\ 
\textbf{PRESERVE DNS} & \checkmark & \checkmark & \checkmark \\ \hline
\end{tabular}
\label{tab:comparison}
\end{table*}
\egroup

\section{Proof-of-concept implementation} 
\label{PrivacyPreservingPassiveDNS}

\subsection{Architecture}

\begin{figure*}[h]
\centering
\includegraphics[width=1\linewidth]{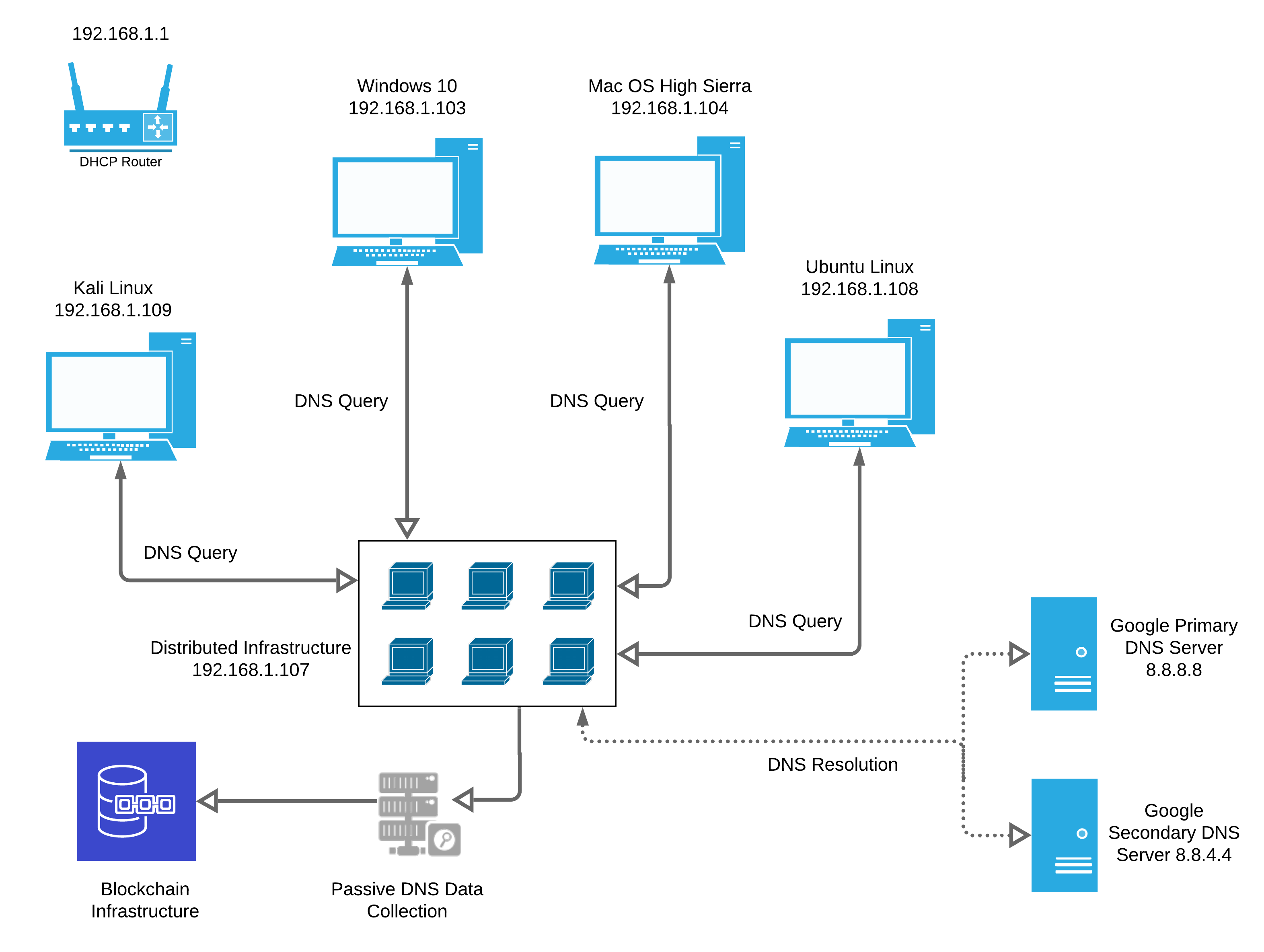}
\caption{\textcolor{black}{PRESERVE DNS proof-of-concept implementation architecture for the test data collection}}
\label{testbed}
\end{figure*}

In order to demonstrate the workings of PRESERVE DNS, and to evaluate its operation and performance, we developed a proof-of-concept implementation, whose architecture is depicted in Figure \ref{testbed}. \textcolor{black}{This proof-of-concept implementation comprises a private data collection that contains the Passive DNS data and is controlled by two authorized \textit{Organizations} with regards to reproducing a Passive DNS infrastructure. Each \textit{Organization} contains two \textit{Peers} that own the blockchain ledger. Each Passive DNS data record consists of ten fields, namely the associated blockchain and record IDs; the domain name; its IP address; the Time-To-Live (TTL); two timestamp fields (in seconds and milliseconds); the number of times that the user visited the domain; the IP address of the end-user; and the server that performed the resolution. The last fields contain personal data and form the private data collection, which is accessible only by the \textit{Peers} of \textit{Organization 1}; others may access only the non-personal data. Note that read and write query times of the \textit{Peers} are measured as part of the evaluation of the proposed infrastructure, reported in the next section.}

As shown in Figure \ref{testbed}, the implementation involves a network of a number of computers that run various operating systems (Microsoft Windows 10, Apple MacOS, Kali Linux and Ubuntu Linux) and use a \textcolor{black}{distributed infrastructure} as a local DNS resolver. \textcolor{black}{Specifically, a Kubernetes pod \cite{vayghan2018deploying} has been configured locally as the DNS resolver of the network. The Kubernetes pod acts as a \textit{host} machine, using its own IP address. The remaining machines use the DNS resolver as their DNS server}, and the local DNS resolver can resolve the DNS queries itself, it can use the ISP's DNS servers, or one of the public DNS servers provided by companies such as Google \cite{Googledns}, Cloudflare \cite{Cloudflaredns} or OpenDNS \cite{Opendns}. In our case, the specified DNS resolver is using Google's public DNS servers for the resolution of the DNS queries, eliminating the chance of a ``bad'' ISP DNS resolver for each DNS query \cite{ager2010comparing}. A passive DNS data collector is needed to collect the DNS queries and responses to store the data in a database for further analysis. This configuration is able to capture queries such as A, AAAA, MX records and the translations between their servers with the domain names. The stored data involves the IPs of the machines that performed the DNS queries and the server that performed the DNS resolution. The DNS resolver passively collects and stores this information in a database, using passivedns from gamelinux \cite{Gamelinux}, and stores the data in JSON format. The system built for the resolution of the DNS queries is using Berkeley Internet Name Domain (BIND) version 9. BIND is the most common implementation of a DNS resolver, and it depends solely on the nameserver it queries \cite{liu2006dns}. The technical specifications of the computers that performed the queries meet the minimum docker container needs. The technical specifications of the distributed infrastructure that hosts the blockchain system are as follows: 6th Generation 2.0GHZ dual-core Intel Core i5 CPU, with 8GB RAM running at 1866 MHz and 256GB PCIe-based flash storage.

\textcolor{black}{This proof-of-concept configuration gives rise to privacy concerns, since IPs can be correlated to the identities of the end-users that visited the sites. However, we emphasize that this configuration is being used only for creating the Passive DNS database; as such its privacy is outside the scope of this paper. The privacy of Passive DNS data collector systems is ensured by PRESERVE DNS that employs a blockchain solution to store the DNS data in an immutable ledger.} 

The blockchain solution should adhere to the following specifications:
\begin{itemize}
    \item Participants of the network should be able to easily query the data stored in the blockchain. 
    \item The queried data should be available only to authorized entities in order to be analysed further for its maliciousness or even to be used as a distributed DNS database protected from various attacks and misuses. 
    \item Consequently, specific data segments on the ledger e.g. IPs of the end-users should be available only to themselves and remain private to all other entities. 
    \item To achieve consensus for storing the data, peers should approve the transaction only for authorized entities who call the corresponding storing procedure. Additionally, for each transaction related to data storage processes, new blocks should be created and added to the ledger, and all the participants should update their ``local'' ledgers to include these new blocks.
\end{itemize}

\subsection{The blockchain}
\begin{figure*}[h]
\centering
\includegraphics[width=1\linewidth]{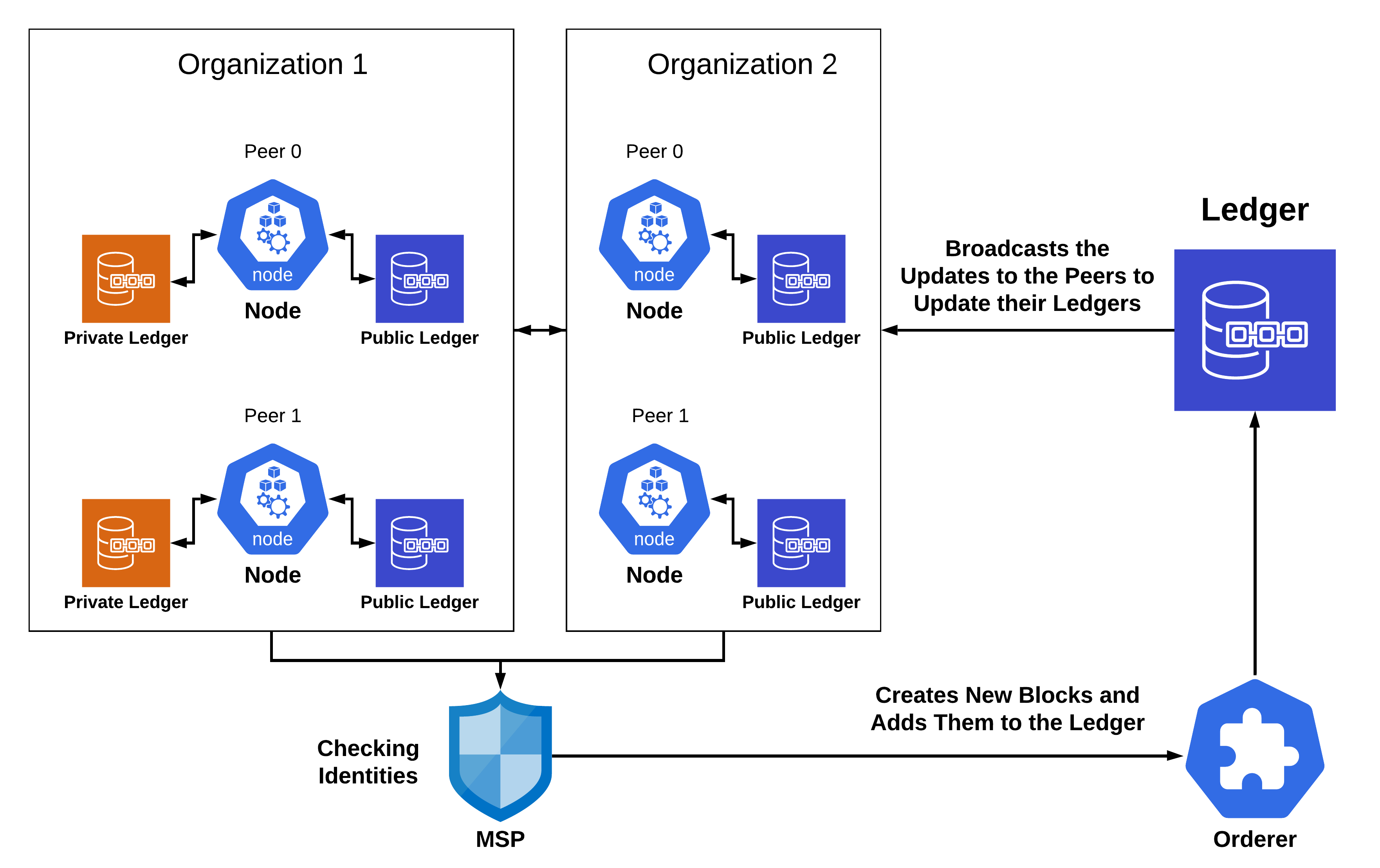}
\caption{\textcolor{black}{Hyperledger Fabric Infrastructure}}
\label{hlftopology}
\end{figure*}

The Hyperledger Fabric platform is used for implementing the blockchain. As illustrated in Figure \ref{hlftopology}, the blockchain infrastructure is composed of two organisations of two peers each; a certificate authority; and an orderer. All the entities of the blockchain are docker containers, authorized for performing their respective purpose according to their identity. The identities issued are X.509 digital certificates signed by the Certificate Authority and checked by the MSP \cite{certificatesFabric}.  During the creation of the blockchain network, the state database that is going be used is defined \cite{androulaki2018hyperledger}; we used \textit{CouchDB}, a complete database available in Hyperledger Fabric that stores data in key-value pairs and also offers rich queries functionality \cite{thakkar2018performance}. The chaincode is written in the Go programming language. 

The first organisation acts as an end-user that stores its passive DNS data collection in the blockchain. Since this data includes personal information (i.e. the IP of the client's machine and the IP of the DNS resolver), they should remain inaccessible by all the other participants of the blockchain, including the second ISP organisation. A private data collection has been used to allow only the first organisation to query for protected data. All the other recipients are able to query only data that they are entitled to access. If an unauthorized peer attempts to query protected data, the query is rejected and an error message is returned.

Hyperledger Fabric provides chaincode APIs to use with command-line (CLI) tools. These chaincode APIs extend the functionalities of the peers and are distinguished in the Init API, Invoke API and Query API classes. Init API is used when initialization or upgrade of the chaincode is executed. Invoke API and Query API are used when storing or reading transactions to the ledger are performed \cite{Hyperledgerdev}.

Peers of the blockchain can store data on the ledger using Hyperledger Fabric's Invoke API. First they have to declare their identity and then use the Invoke API with the corresponding storing function and the arguments in JSON format to send each transaction to the orderer. The orderer receives the data and performs the requested storing function. In case of success, this procedure will create a new block on the ledger and will send an update signal to each of the peers to update their ledgers.

To receive data from the ledger, peers use the Query API. They declare their identity and then use the corresponding query function with the arguments in JSON format to send the transaction to the orderer. The orderer receives the transaction and subsequently displays only those data to the recipients that they are allowed to access according to the defined query function and the private data collection configuration. The specified identity functionality enables a peer to query only specific blocks. Some data, such as the IPs of each end-user, must remain private and available only to them. 

Some fundamental differences exist between the ``traditional'' public and the ``permissioned'' blockchain infrastructures as discussed in Sections \ref{blockchain} and \ref{hlftopologysection}. One of these is the consensus mechanism, that in Hyperledger Fabric can be configured. The consensus mechanism used in the proof-of-concept implementation of PRESERVE DNS requires at least one peer from any organization to accept a transaction, in order for the transaction to be considered valid. Another major difference is the time taken for transaction processing \cite{sompolinsky2015secure}. Bitcoin requires approximately ten minutes per new block that contains a few transactions. Compared to that, Hyperledger Fabric can process a few thousands transactions per second \cite{androulaki2018hyperledger}, while maintaining the promoted privacy and security. A similar feature in-context in ``traditional'' blockchains with Hyperledger Fabric is the Peers entity that could be compared to ``miners'' or full nodes \cite{bashir2017mastering}. 

\section{Evaluation}
\label{evaluation}
\subsection{Security evaluation - DNS attacks}
\label{seceval}
Common DNS attacks are DNS DDoS \cite{kambourakis2007detecting}, DNS fast-flux \cite{ranjan2012detecting} and DNS cache poisoning \cite{schuba1993addressing}; PRESERVE DNS can thwart these attacks as follows: 
\begin{itemize}
    \item The proof-of-concept implementation of PRESERVE DNS described in the previous section contains only one orderer. A potential DDoS attack against the orderer container may result in particular writes to the ledger to be blocked. However, in a production environment, this attack can be prevented using more ordering services under the same Kafka cluster. When one orderer fails, the Kafka cluster assigns another orderer to complete the transaction.
    \item In a fast-flux DNS attack a malicious actor uses short-timed time-to-live (TTL) records to change legitimate to malicious servers under the same hostname. PRESERVE DNS thwarts Fast-flux DNS attacks, since the administrator of the blockchain configures the TTL of the ledger's blocks.
    \item PRESERVE DNS is able to thwart DNS cache poisoning attacks, but only if it is being used as the local DNS database in the system, as in the case of our proof-of-concept implementation. This means that the local DNS resolver should query PRESERVE DNS for every DNS query instead of using the local DNS cache first. A potential solution to this issue in a production environment is to continually update the local DNS cache with the data from the blockchain using a scheduled job. 
\end{itemize}

\subsection{Security evaluation - Blockchain attacks}
Blockchain is an immutable ledger, and the data stored cannot be manipulated by malicious actors. Each transaction needs to be authorized by the policy, and unauthorized requests are rejected automatically. All the functions and security mechanisms of the blockchain are included in the chaincode that is installed in each participant. A collection configuration is developed to advise the orderer about the state of the stored data, the time of their availability until they purge, and all entities that have access to them. 

Each peer is obliged to prove its identity to the orderer before being allowed to perform a transaction. According to the configured policy, the store and query transactions are restricted to peers which are not included in the policy, \textcolor{black}{thus preserving the privacy of the stored data. These fundamental principles eliminate the possibility of an unauthorized, malicious actor to store arbitrary data to the ledger. Furthermore, a malicious actor is overall unable to query data. The personal data can be queried only by specified entities and the remaining data are available only to participants.}

DDoS attacks against the blockchain are thwarted as well. Hyperledger Fabric is using docker containers to act as peers of the blockchain. Each peer has the whole ledger stored, including the history of each transaction. PRESERVE DNS is composed of Docker containers acting as the participants of the network. This means that when a Docker container fails the query is passed on to another active peer. When the failed peer recovers, it initiates the \textit{gossip protocol} to update its ledger with missing entries, from the rest of the peers. For a successful DDoS attack to take place against PRESERVE DNS, all the peers should be successfully attacked at the same time; this is highly improbable and practically impossible. Another solution to this problem is to develop the docker containers inside a Kubernetes cluster. In case of a container failure, Kubernetes is able to restart it or create an identical twin to operate as the failed one.

Attacks such as DNS amplification attacks and some zero-day DNS attacks against the blockchain infrastructure are not possible, because to launch them data in the ledger would need to be altered. PRESERVE DNS is operationally resilient, since it is composed of distributed peer nodes and organisations that have identical ledgers. 

Further, infrastructure continuity in PRESERVE DNS is ensured as long as at least one peer node is operating normally \cite{Deloitteblockchain}.  Accordingly, PRESERVE DNS has no single point of failure. 

A significant issue for permissioned blockchains such as the Hyperledger Fabric is the human factor, considered to be the weakest link in any system's security chain. Key management is a crucial part of the blockchain infrastructure's security. Potential theft of a legitimate user's identity by a malicious actor can lead to unauthorized queries and arbitrary storage of data in the ledger. Additionally, the chaincode is compiled and can run completely autonomously, but it is coded by humans. thus, it is possible that the code contains bugs that are discovered late in the infrastructure lifecycle \cite{androulaki2018hyperledger}. 

Despite the enhanced security that the blockchain technology enjoys, it is still young, and new kinds of attacks are most likely to be discovered in the future. One potential source of threats is the advancement of quantum computing, against which most of the existing cryptographic techniques are defenseless. Thus, whenever possible, it is recommended to use a quantum-robust algorithm if its overhead on the performance of the system is tolerable \cite{Microsoftblockchain}.

\subsection{Performance evaluation} 
\textcolor{black}{The last part of the evaluation compares the performance of PRESERVE DNS against an alternative blockchain solution, namely Blockstack, and against a traditional database that offers column level privacy, such as PostgreSQL.} 
Blockstack uses by default the Gaia decentralised storage system \cite{ali2017blockstack}. Even if it does not offer the enhanced privacy of PRESERVE DNS, it is a promising alternative to the current DNS infrastructure. On the other hand, the \textcolor{black}{PostgreSQL} database is the most popular option for production environments because it is a database server that offers extra functionalities, such as remote connections \cite{momjian2001postgresql}. Additionally, the \textcolor{black}{PostgreSQL} database server can run isolated in a Docker container \cite{boettiger2015introduction}. 

\textcolor{black}{As already discussed in Section~\ref{Relatedwork}, the DNSTSM system \cite{yu2020dnstsm} is very similar to ours. However, being based on an older version of Hyperledger Fabric, it is unable to create a complete privacy-preserving infrastructure. Hence, we did not compare its performance to that of PRESERVE DNS, and we limited ourselves to a comparison of security levels, presented in Table~\ref{tab:comparison}.} 

\begin{figure}[h]
\centering
\includegraphics[width=0.8\linewidth]{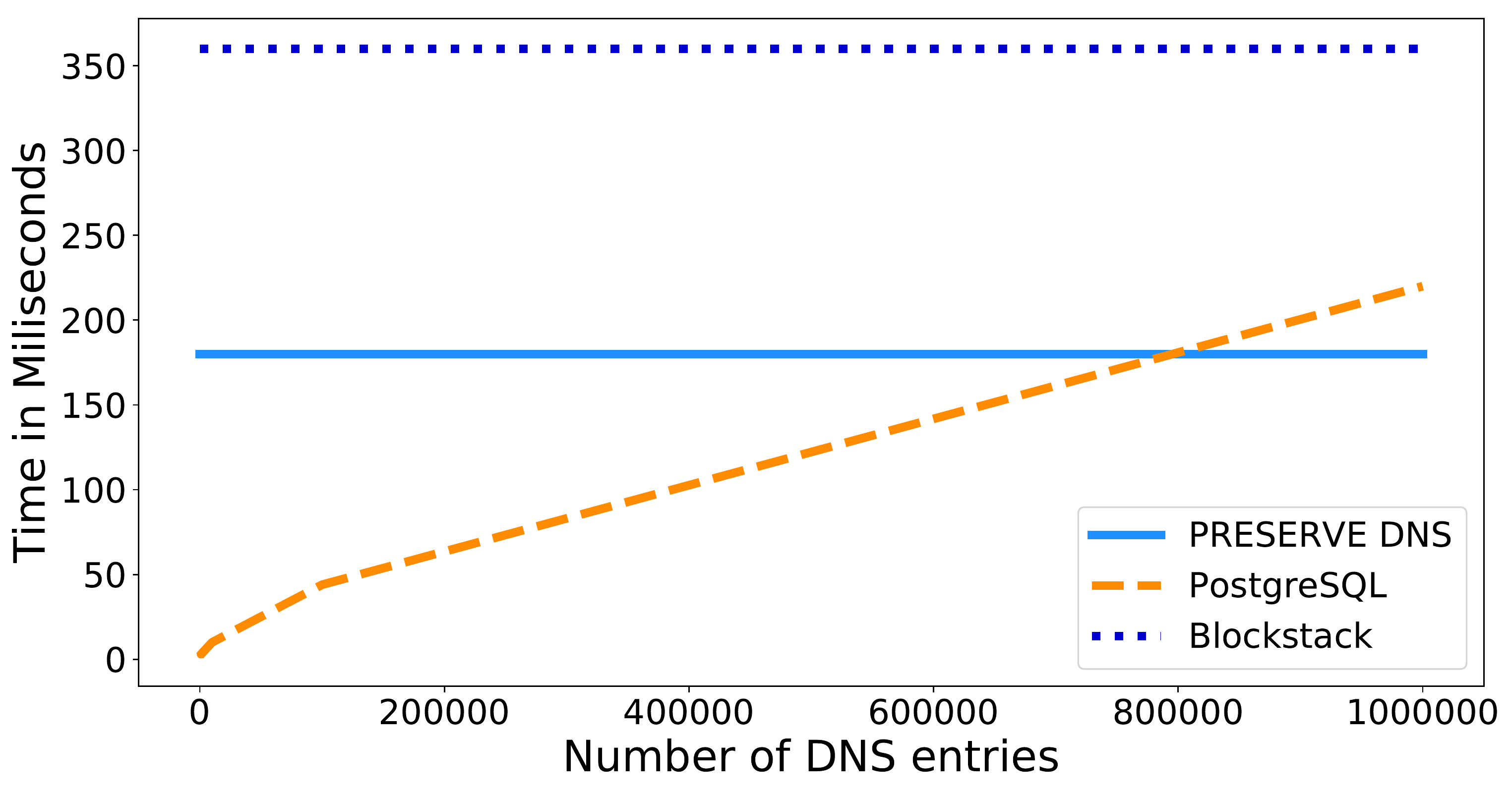}
\caption{\textcolor{black}{Read Data transactions overhead}}
\label{perf_fig}
\end{figure}

The time taken to perform a (a) Read data and (b) Write Data transaction in each of the three alternatives (PRESERVE DNS, \textcolor{black}{PostgreSQL} database and Blockstack's Gaia decentralised storage) for various numbers of DNS entries \textcolor{black}{(10, 1000, 10.000, 100.000, 1.000.000)} is used as the performance metric. The results are depicted in Table~\ref{tab:benchmarks} and in Figure~\ref{perf_fig}. As illustrated in Figure~\ref{perf_fig} the \textcolor{black}{PostgreSQL} database query time (or query overhead) is less for a small number of DNS Entries but raises linearly with each additional DNS entry. Data stored in Blockstack's Gaia decentralised storage is in the form of key-value pairs and is stored off-chain. On the other hand, the query time in PRESERVE DNS is unchanged, since it queries indexed items stored in a distributed ledger. The benefits of using a distributed solution, such as PRESERVE DNS are even greater in production environments because the passive DNS data analysis consists of millions of DNS entries and scaling is necessary. \textcolor{black}{All comparisons were done over the same architecture, illustrated in Section~\ref{PrivacyPreservingPassiveDNS}; they were executed independently and in isolation, to eliminate bias and ensure the accuracy of the results.}

\bgroup
\def\arraystretch{1.3}
\begin{table*}[ht]
\centering
\caption{\textcolor{black}{Read Data/Write Data transaction time in milliseconds (ms) per number of DNS entries}} 
\begin{tabular}{|c|c|c|c|c|c|c|}
\hline
\multicolumn{2}{|c|}{\textbf{Number of DNS Entries}} & \textbf{10} & \textbf{1000} & \textbf{10,000} & \textbf{100,000} & \textbf{1,000,000} \\ [0.5ex] \hline
\multirow{2}{*}{\textbf{\textcolor{black}{PRESERVE DNS}}} & \textit{Read Data} & 180ms & 180ms & 180ms & 180ms & 180ms \\
 & \textit{Write Data} & 230ms & 230ms & 230ms & 230ms & 230ms \\ \hline
\multirow{2}{*}{\textbf{\textcolor{black}{PostgreSQL Database}}} & \textit{Read Data} & 2ms & 3ms & 10ms & 44ms & 220ms \\
 & \textit{Write Data} & 4ms & 5ms & 6ms & 9ms & 11ms \\ \hline
 \multirow{2}{*}{\textbf{Blockstack Ali et al. \cite{ali2017blockstack}}} & \textit{Read Data} & 360ms & 360ms & 360ms & 360ms & 360ms \\
 & \textit{Write Data} & 530ms & 530ms & 530ms & 530ms & 530ms \\ \hline
\end{tabular}
\label{tab:benchmarks}
\end{table*}
\egroup

\begin{figure}[h!]
    \centering
    \subfloat[\textcolor{black}{Read queries workflow on 1.000 DNS Entries}]{{\includegraphics[width=7cm]{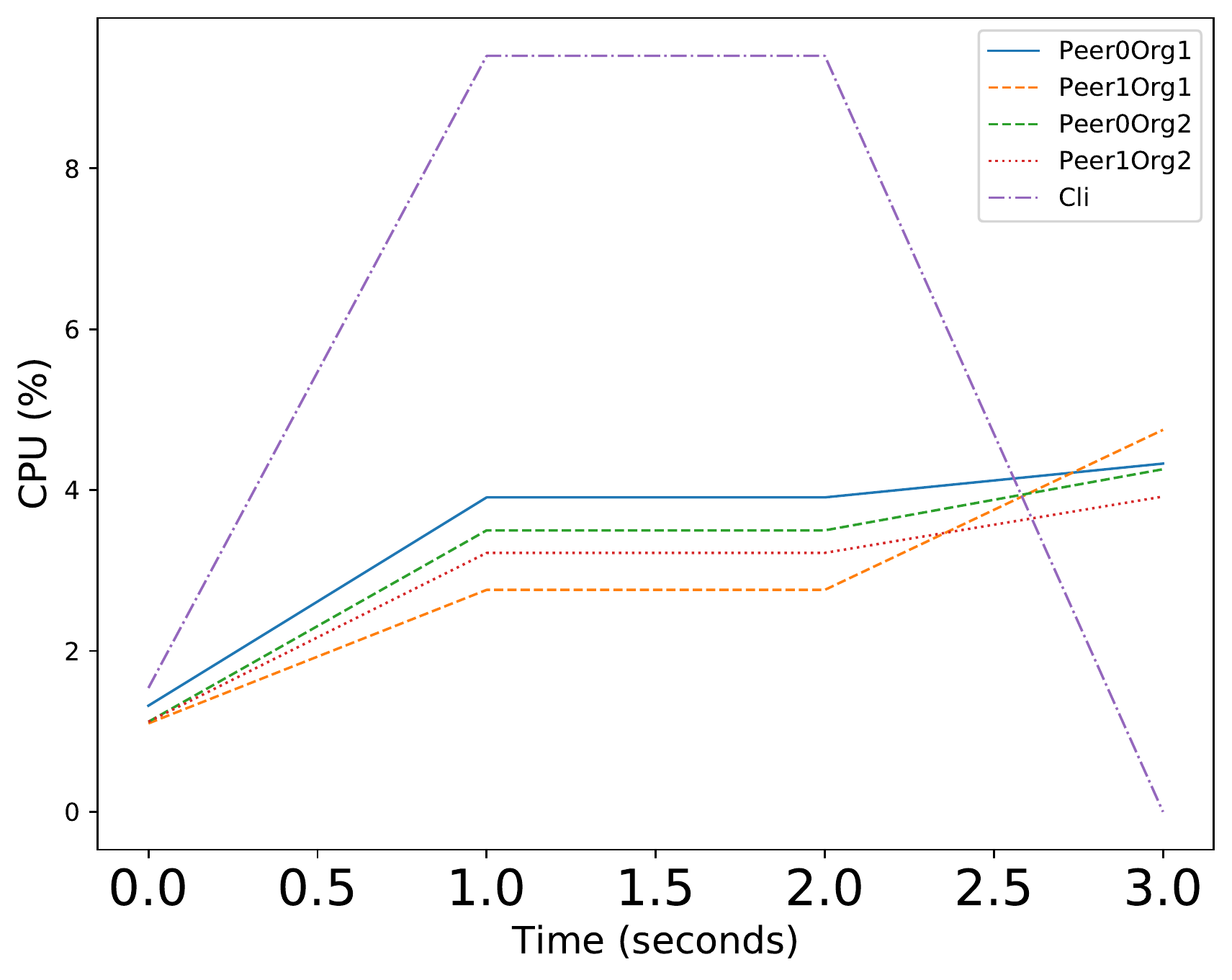} }}%
    \qquad
    \subfloat[\textcolor{black}{Write queries workflow on 1.000 DNS Entries}]{{\includegraphics[width=7cm]{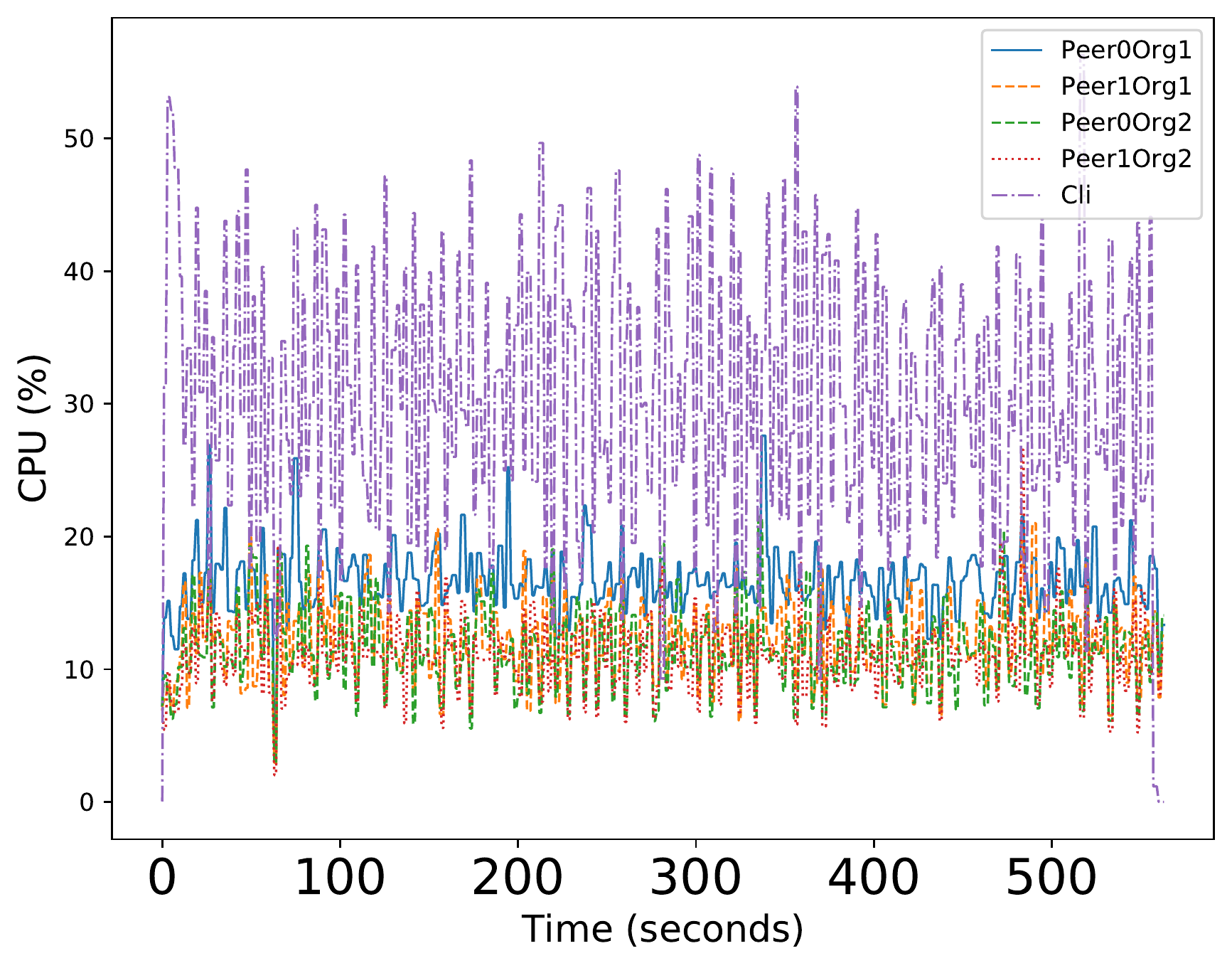} }}
    \qquad 
    \subfloat[\textcolor{black}{Read queries workflow on 10.000 DNS Entries}]{{\includegraphics[width=7cm]{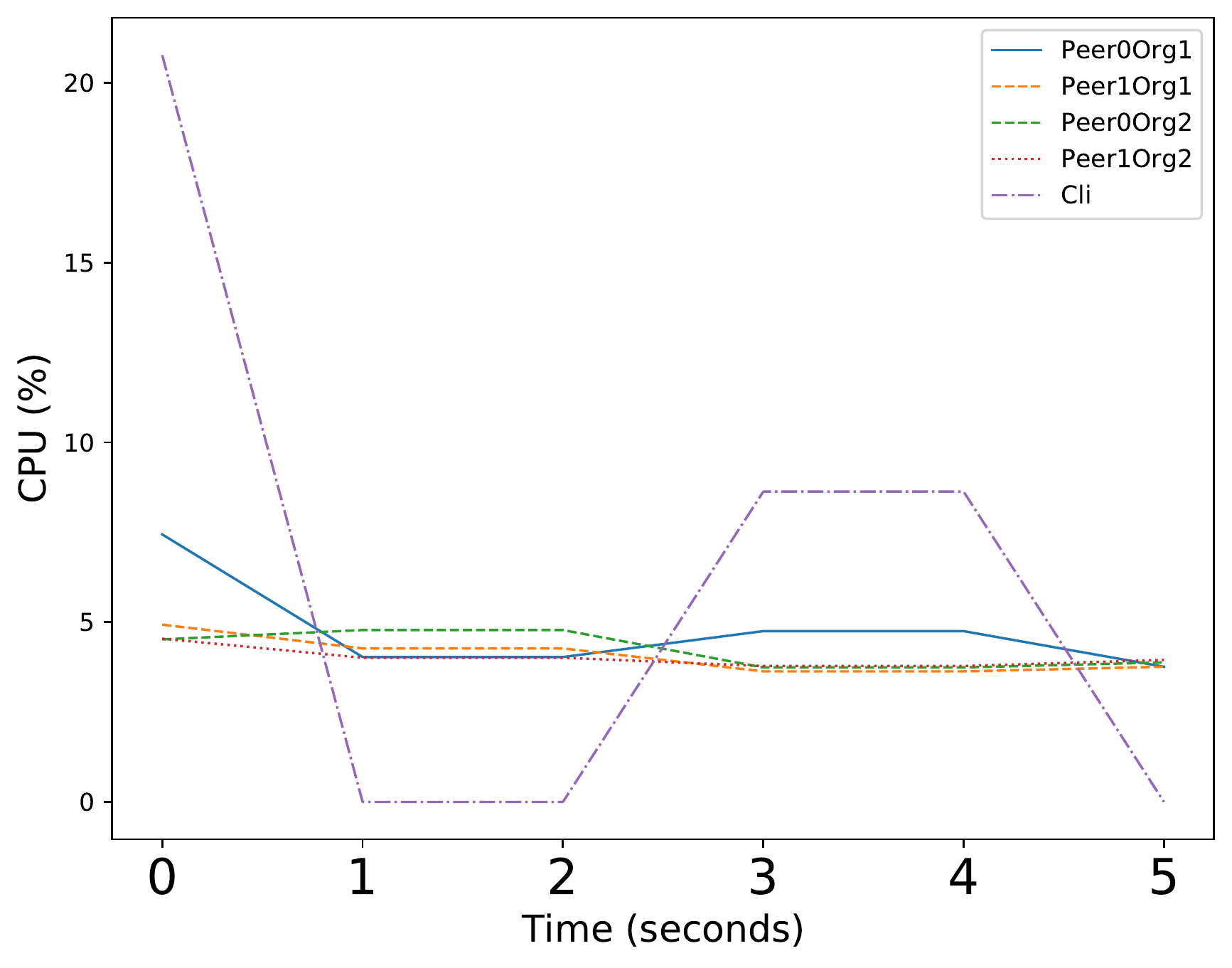} }}%
    \qquad
    \subfloat[\textcolor{black}{Write queries workflow on 10.000 DNS Entries}]{{\includegraphics[width=7cm]{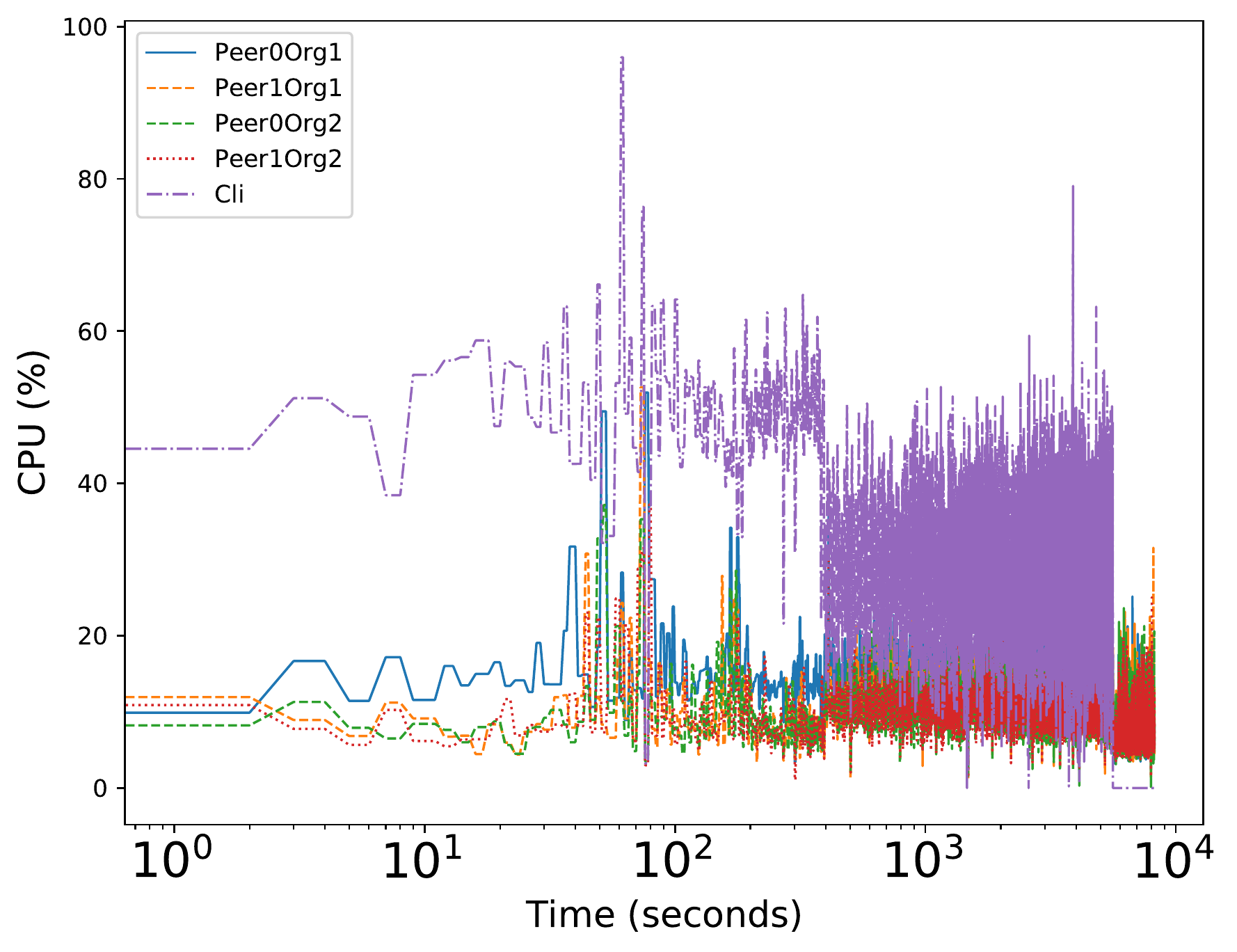} }}
    \qquad
    \subfloat[\textcolor{black}{Read queries workflow on 100.000 DNS Entries}]{{\includegraphics[width=7cm]{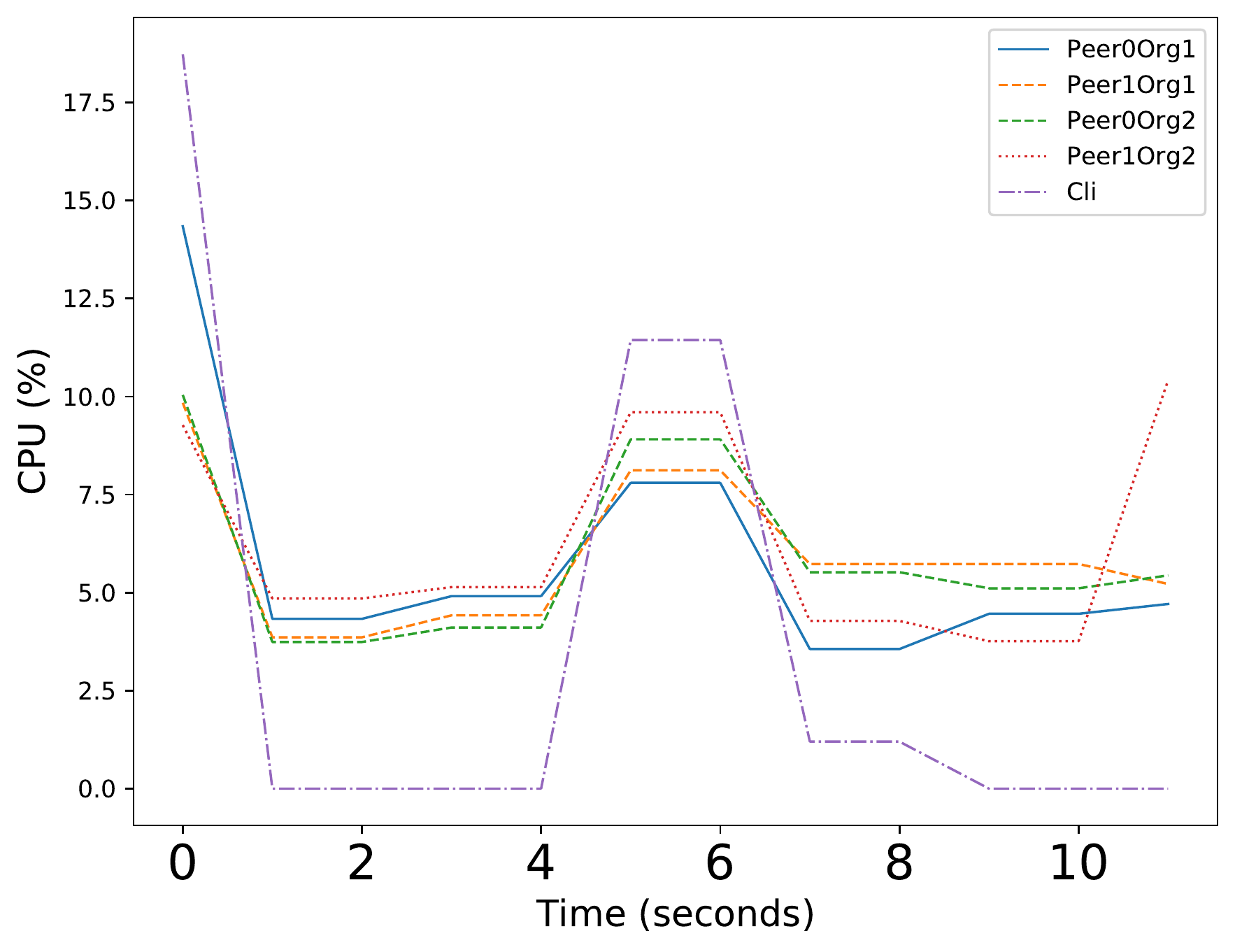} }}
    \qquad 
    \subfloat[\textcolor{black}{Write queries workflow on 100.000 DNS Entries}]{{\includegraphics[width=7cm]{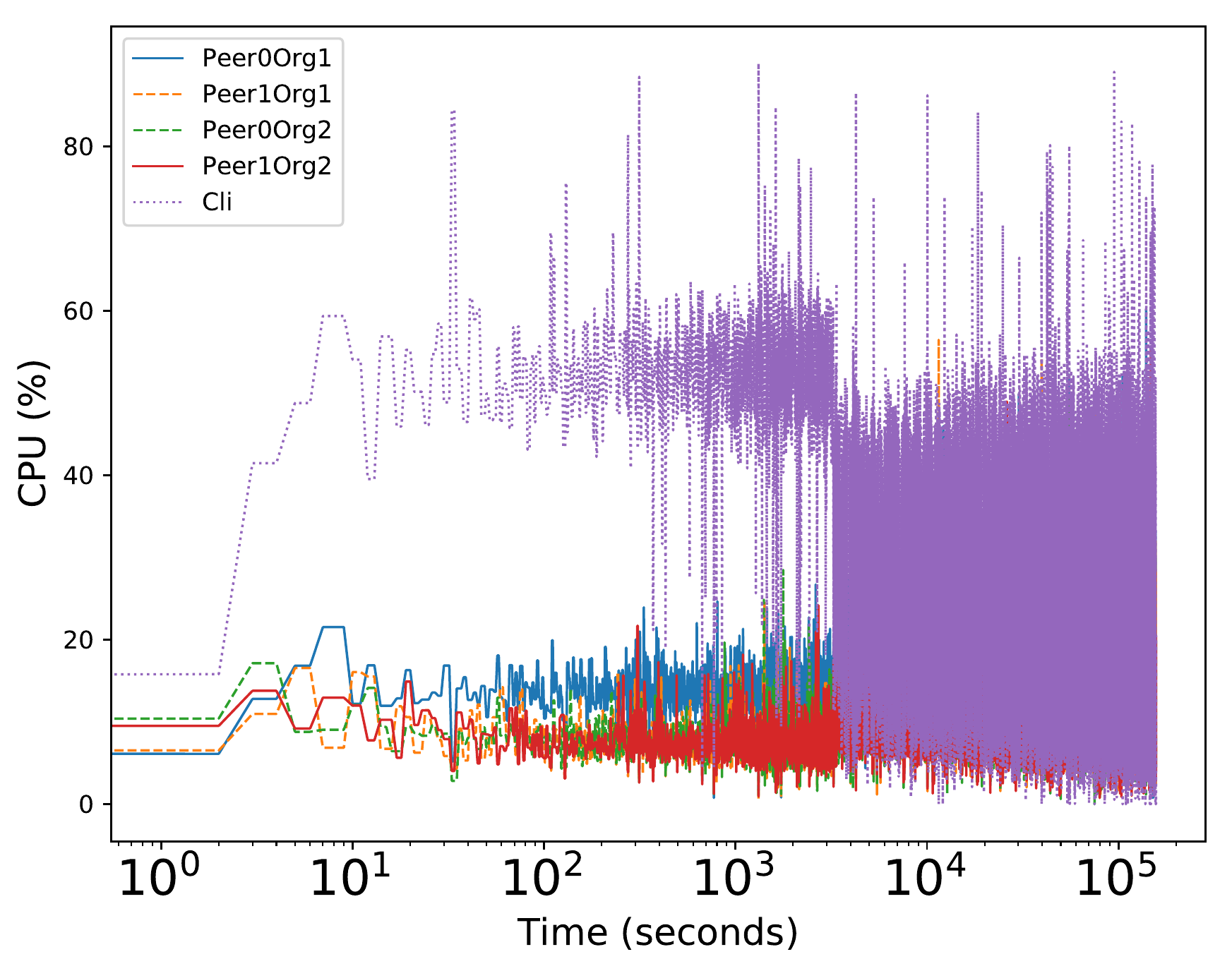} }}
    \caption{\textcolor{black}{CPU Usage (\%) of Nodes during workflow}}
    \label{fig:cpuplots}
\end{figure}

\textcolor{black}{Furthermore, we benchmarked the CPU and Memory performance of our proposed solution. Figure~\ref{fig:cpuplots} shows that the CPU usage of the blockchain nodes during Read queries workflow is low (5\%-10\%) over a varying number of DNS entries (1.000, 10.000, 100.000). Additionally, the Write queries workflow follows a similar CPU usage pattern (< 20\%). The blockchain nodes in the form of Docker containers are distinguished in the first and second Peers (Peer0, Peer1) of Organization 1 and Organization 2 (Org1, Org2). Moreover, the last \textit{CLI} container is the container that is being used by Hyperledger Fabric for the command-line interface usability \cite{CliHyperledger}. In Figures~\ref{fig:cpuplots} (d) and (e), the \textit{CLI} container's CPU usage fluctuates rapidly and conceals the CPU usage of the blockchain nodes. This is reasonable as the \textit{CLI} container is being directly used by the authors for each transaction. Furthermore, it is also noteworthy that the CPU usage of the blockchain nodes is less than 20\%.}

\begin{figure*}[h!]
    \centering
    \subfloat[\textcolor{black}{First Peer (Peer0) of Organization 1 (Org1) container workflow}]{{\includegraphics[width=7cm]{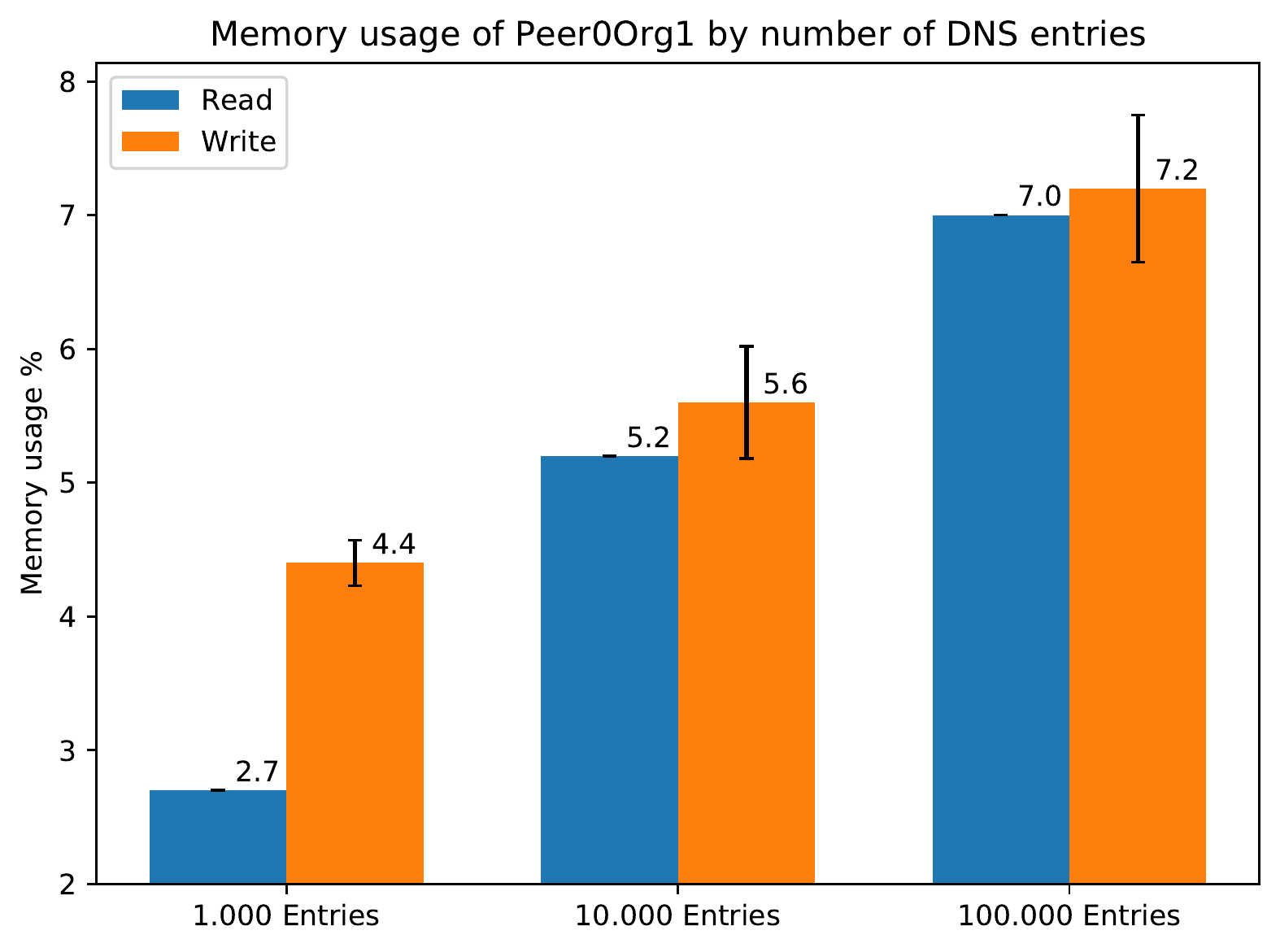} }}
    \qquad 
    \subfloat[\textcolor{black}{Second peer (Peer1) of Organization 1 (Org1) container workflow}]{{\includegraphics[width=7cm]{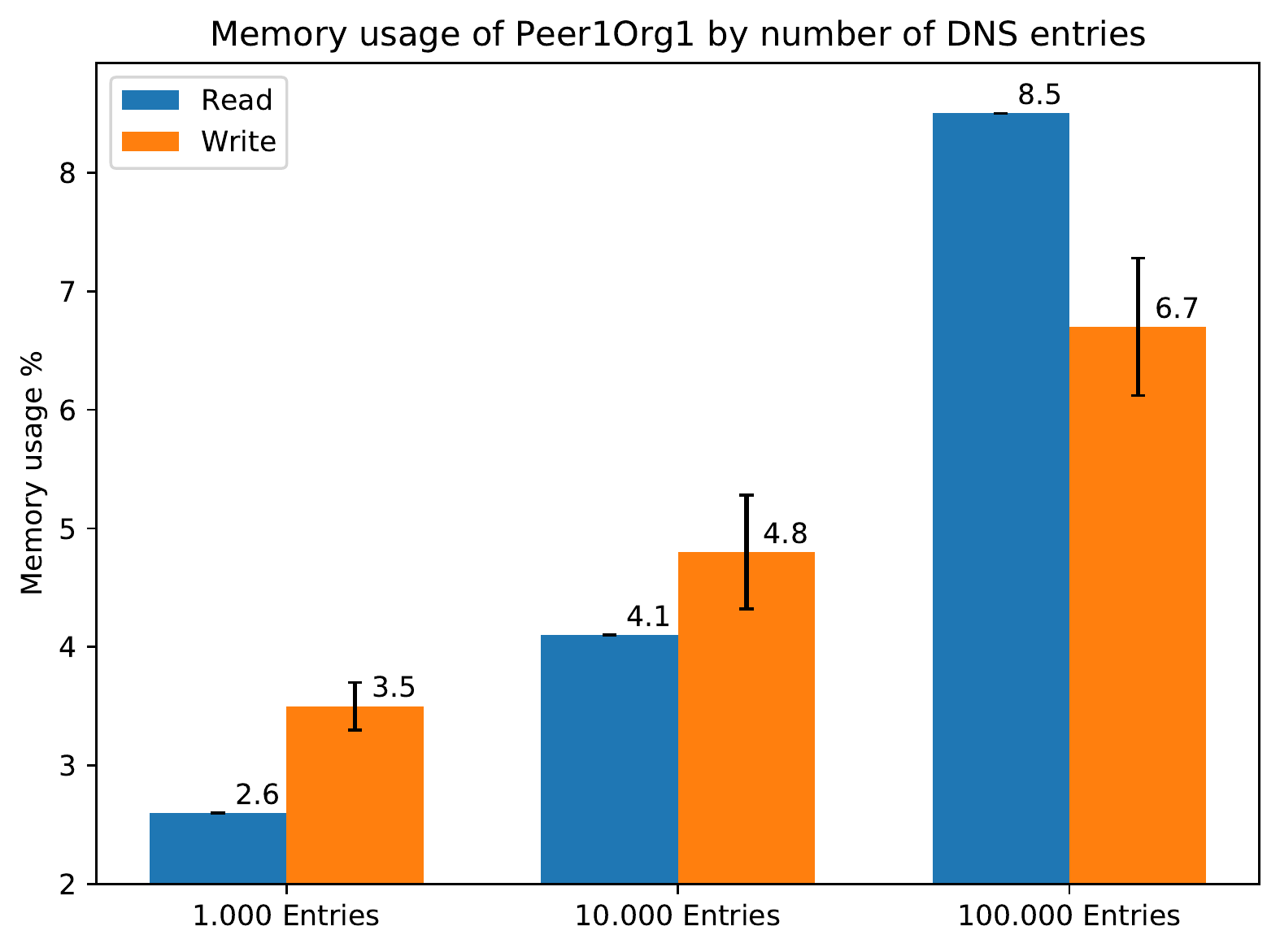} }}
    \qquad
    \subfloat[\textcolor{black}{First Peer (Peer0) of Organization 2 (Org2) container workflow}]{{\includegraphics[width=7cm]{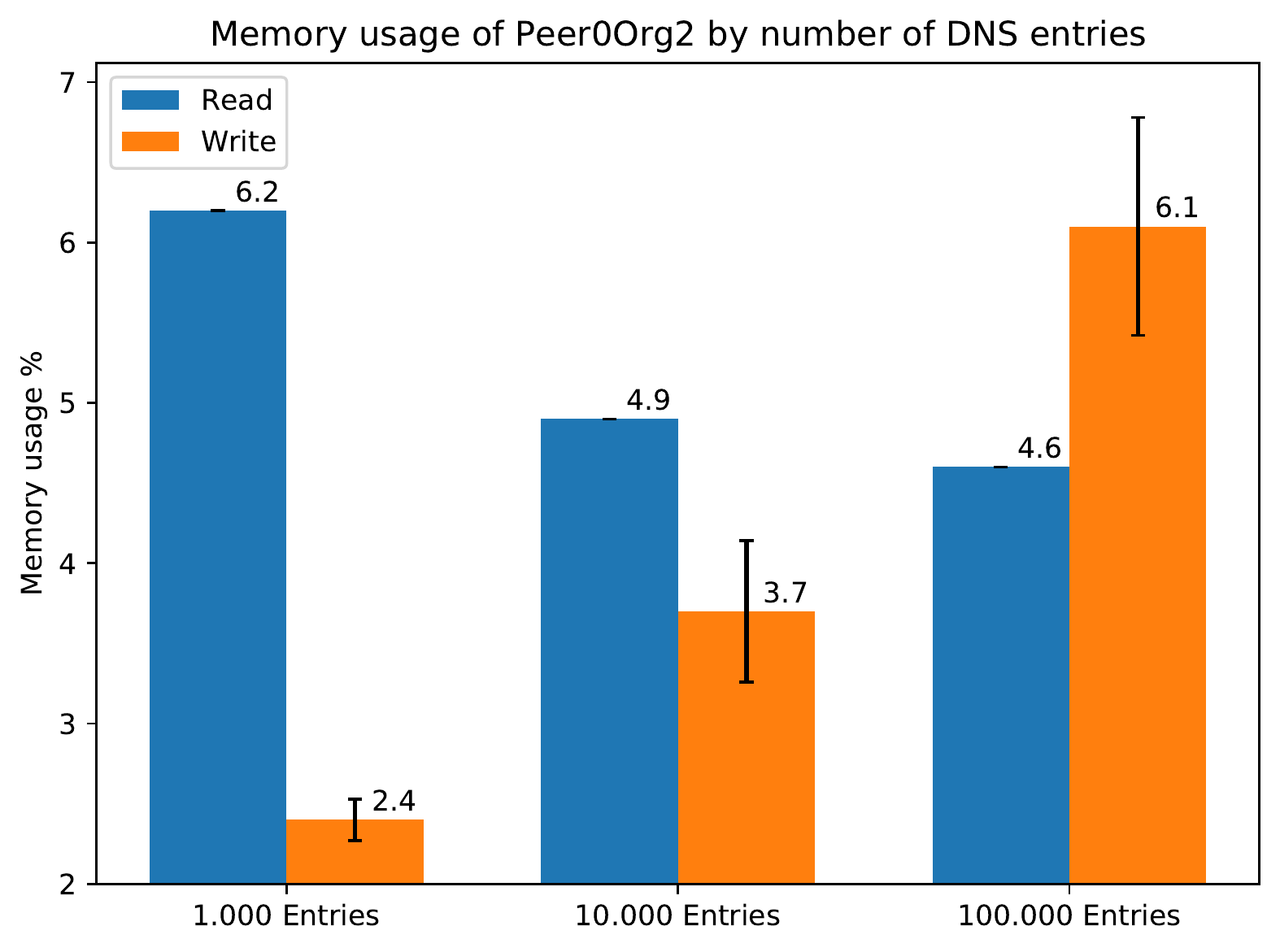} }}
    \qquad
    \subfloat[\textcolor{black}{Second Peer (Peer1) of Organization 2 (Org2) container workflow}]{{\includegraphics[width=7cm]{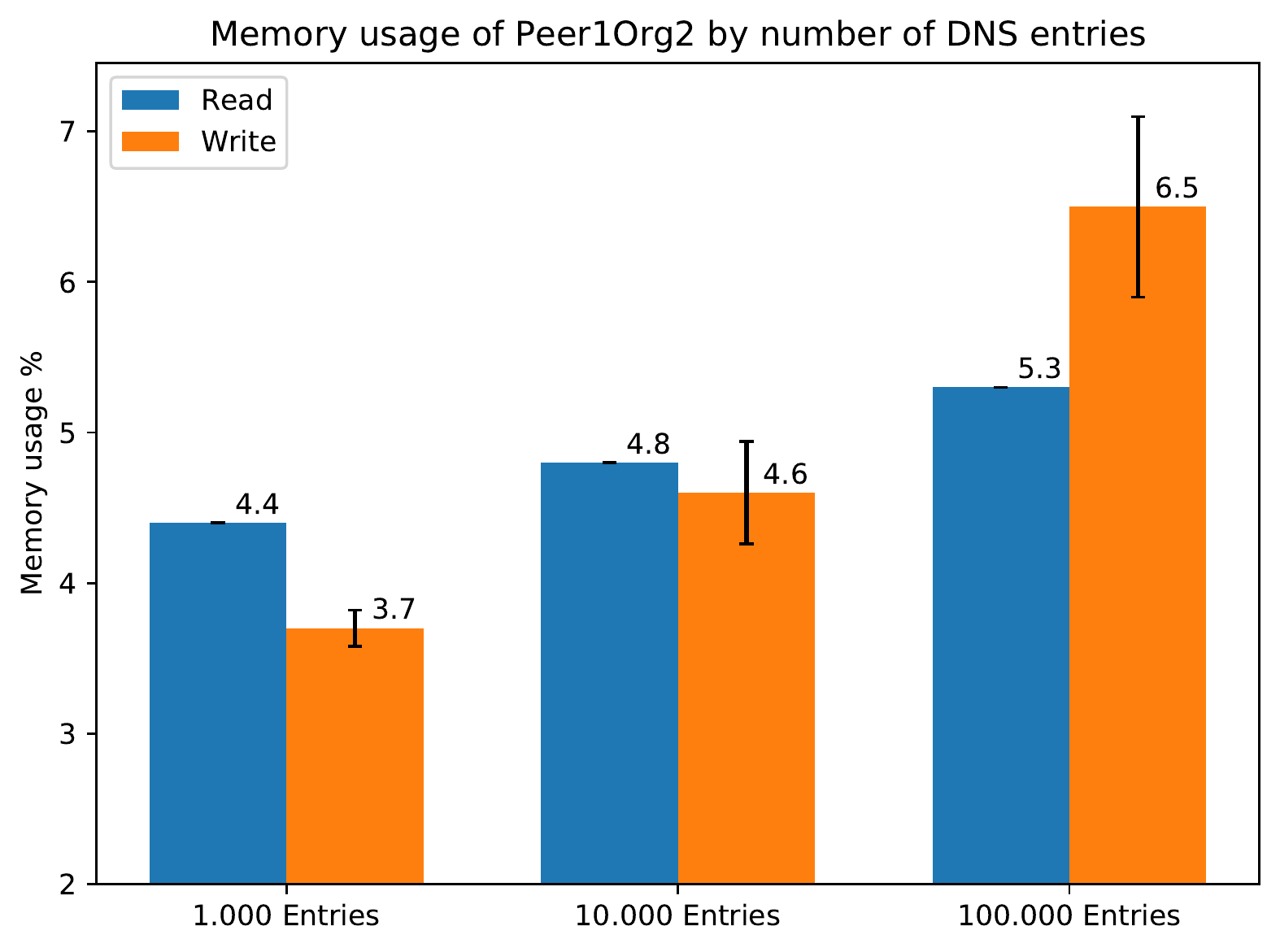} }}
    \caption{\textcolor{black}{Memory Usage (\%) of Nodes during workflow on 1.000, 10.000 and 100.000 DNS Entries}}
    \label{fig:memoryplots}
\end{figure*}

\textcolor{black}{Figure~\ref{fig:memoryplots} depicts the average memory usage of each blockchain node, in addition to the minimum and maximum values scored. The memory usage for each Read and Write query measured over a varying number of DNS entries (1.000, 10.000, 100.000) is considerably low. In this figure the naming of the docker containers follows the convention described above. Contrary to Figure~\ref{fig:cpuplots}, the memory usage of the \textit{CLI} container does not fluctuate significantly.}

\section{Conclusions and Future Work} 
\label{ConclusionsandFutureWork}

DNS is the Internet's phone book. Its services are and will continue to be invaluable for years to come. Its main setback is that it was built without strong consideration of its security, thus allowing malicious parties to abuse it. Part of the defence against such abuse is the processing of DNS resolution data, collected by passive DNS, towards identifying malicious actors such as e.g. malicious domain names. However, this process involves collecting and processing personal end-user data. As the whole world becomes data centric, securing such data becomes paramount, hence the need for privacy-preserving passive DNS. 

\textcolor{black}{We proposed PRESERVE DNS, an environment that can be used for passive DNS data analysis whilst ensuring end-user privacy. We provided a proof-of-concept implementation and we evaluated the performance of PRESERVE DNS against another blockchain solution, namely Blockstack, and against a traditional database that offers column level privacy, such as PostgreSQL. PRESERVE DNS was found to be resilient to various attacks and to impose less query overhead in high volumes of data than that of existing alternatives. In our proof-of-concept implementation, PRESERVE DNS achieves a read query time of 180 milliseconds in every tested number of DNS entries. Similarly, in every tested number of DNS entries, Blockstack achieves a flat read query time of 360 milliseconds. In contrast, the traditional PostgreSQL database read query time increases over higher volumes of DNS entries, achieving 220 milliseconds in 1.000.000 entries.}

Existing techniques for analysing passive DNS data, such as Notos \cite{antonakakis2010building}, EXPOSURE \cite{bilge2011exposure}, and \cite{khalil2016discovering} can use PRESERVE DNS to achieve the same functionality without violating the privacy of the end-users. PRESERVE DNS can also be set-up online, to provide further services to end-users. It can be used as a public DNS resolver that also passively collects the queries and responses, for malicious domain name detection, transparently to the end-users. PRESERVE DNS can be part of an ISP's or a TLD's infrastructure, or it can be used as part of a passive DNS data analysis system. In the latter option, users that store their DNS data to the system for further analysis can be sure that their privacy is not violated and only themselves may query their personal data. PRESERVE DNS achieves this trustworthiness without the need for users to trust another third party or the system itself.

As future work, we plan to extend the functionality of PRESERVE DNS, and to produce a cloud-oriented implementation, that would facilitate its adoption by ISPs. The advantages of placing the services of PRESERVE DNS in the cloud or on a cloud service provider could multiply if it is placed on a Kubernetes cluster. This would provide operational continuity with semi-infinitely scalability and expandability. Since PRESERVE DNS uses the Hyperledger Fabric platform that consists of Docker containers, making a transition to a Kubernetes cluster is a viable and realistic option.

\authorcontributions{All authors contributed in the conceptualization and methodology of the manuscript; P.P. performed the data preparation; P.P. and N.P. contributed in writing; W.J.B., O.L., S.K. reviewed and edited the manuscript. All authors have read and agreed to the published version of the manuscript.}

\funding{This research received no external funding.}

\conflictsofinterest{The authors declare no conflict of interest.} 

\reftitle{References}
\externalbibliography{yes}
\bibliography{references}

\begin{thebibliography}{-------}
\providecommand{\natexlab}[1]{#1}

\bibitem[Mockapetris(1987)]{mockapetris1987domain}
Mockapetris, P.V.
\newblock Domain names-concepts and facilities,  1987.

\bibitem[Antonakakis \em{et~al.}(2017)Antonakakis, April, Bailey, Bernhard,
  Bursztein, Cochran, Durumeric, Halderman, Invernizzi, Kallitsis,
  et~al.]{antonakakis2017understanding}
Antonakakis, M.; April, T.; Bailey, M.; Bernhard, M.; Bursztein, E.; Cochran,
  J.; Durumeric, Z.; Halderman, J.A.; Invernizzi, L.; Kallitsis, M.; others.
\newblock Understanding the mirai botnet.
\newblock  26th $\{$USENIX$\}$ Security Symposium ($\{$USENIX$\}$ Security 17),
   2017, pp. 1093--1110.

\bibitem[Vissers \em{et~al.}(2015)Vissers, Joosen, and
  Nikiforakis]{vissers2015parking}
Vissers, T.; Joosen, W.; Nikiforakis, N.
\newblock Parking sensors: Analyzing and detecting parked domains.
\newblock  Proceedings of the 22nd Network and Distributed System Security
  Symposium (NDSS 2015). Internet Society,  2015, pp. 53--53.

\bibitem[Stout and McDowell(2012)]{stout2012system}
Stout, B.; McDowell, K.
\newblock System and method for combating cybersquatting,  2012.
\newblock US Patent 8,285,830.

\bibitem[Weimer(2005)]{weimer2005passive}
Weimer, F.
\newblock Passive DNS replication.
\newblock  FIRST conference on computer security incident,  2005, p.~98.

\bibitem[GDP(2016)]{GDPRip}
{Patrick Breyer v Bundesrepublik Deutschland Case C‑582/14
  ECLI:EU:C:2016:779},  2016.

\bibitem[Spring and Huth(2012)]{spring2012impact}
Spring, J.M.; Huth, C.L.
\newblock The impact of passive dns collection on end-user privacy.
\newblock {\em Securing and Trusting Internet Names} {\bf 2012}.

\bibitem[{Google}(2018)]{Googledns}
{Google}.
\newblock {Google Public DNS},  2018.

\bibitem[{Cloudflare}(2018)]{Cloudflaredns}
{Cloudflare}.
\newblock {What is 1.1.1.1?},  2018.

\bibitem[{OpenDNS}(2006)]{Opendns}
{OpenDNS}.
\newblock {OpenDNS},  2006.

\bibitem[Federrath \em{et~al.}(2011)Federrath, Fuchs, Herrmann, and
  Piosecny]{federrath2011privacy}
Federrath, H.; Fuchs, K.P.; Herrmann, D.; Piosecny, C.
\newblock Privacy-preserving DNS: analysis of broadcast, range queries and
  mix-based protection methods.
\newblock  European Symposium on Research in Computer Security. Springer,
  2011, pp. 665--683.

\bibitem[Khalil \em{et~al.}(2016)Khalil, Yu, and Guan]{khalil2016discovering}
Khalil, I.; Yu, T.; Guan, B.
\newblock Discovering malicious domains through passive DNS data graph
  analysis.
\newblock  Proceedings of the 11th ACM on Asia Conference on Computer and
  Communications Security. ACM,  2016, pp. 663--674.

\bibitem[Gasser \em{et~al.}(2018)Gasser, Hof, Helm, Korczynski, Holz, and
  Carle]{gasser2018log}
Gasser, O.; Hof, B.; Helm, M.; Korczynski, M.; Holz, R.; Carle, G.
\newblock In Log We Trust: Revealing Poor Security Practices with Certificate
  Transparency Logs and Internet Measurements.
\newblock  International Conference on Passive and Active Network Measurement.
  Springer,  2018, pp. 173--185.

\bibitem[{Farsight Security}(2010)]{Farsightdnsdb}
{Farsight Security}.
\newblock {DNSDB},  2010.

\bibitem[{VirusTotal}(2013)]{Virustotalpassivedns}
{VirusTotal}.
\newblock {VirusTotal Passive DNS replication},  2013.

\bibitem[James(2005)]{james2005phishing}
James, L.
\newblock {\em Phishing exposed}; Elsevier,  2005.

\bibitem[Yoon \em{et~al.}(2019)Yoon, Kim, Kim, Shin, and
  Son]{yoon2019doppelgangers}
Yoon, C.; Kim, K.; Kim, Y.; Shin, S.; Son, S.
\newblock Doppelg{\"a}ngers on the Dark Web: A Large-scale Assessment on
  Phishing Hidden Web Services.
\newblock  The World Wide Web Conference,  2019, pp. 2225--2235.

\bibitem[Riederer \em{et~al.}(2011)Riederer, Erramilli, Chaintreau,
  Krishnamurthy, and Rodriguez]{riederer2011sale}
Riederer, C.; Erramilli, V.; Chaintreau, A.; Krishnamurthy, B.; Rodriguez, P.
\newblock For sale: your data: by: you.
\newblock  Proceedings of the 10th ACM WORKSHOP on Hot Topics in Networks,
  2011, pp. 1--6.

\bibitem[Bortzmeyer(2015)]{bortzmeyer2015dns}
Bortzmeyer, S.
\newblock DNS privacy considerations.
\newblock {\em Work in Progress, draft-ietf-dprive-problem-statement-06} {\bf
  2015}, {\em 1}.

\bibitem[Wang \em{et~al.}(2019)Wang, Hoang, Hu, Xiong, Niyato, Wang, Wen, and
  Kim]{wang2019survey}
Wang, W.; Hoang, D.T.; Hu, P.; Xiong, Z.; Niyato, D.; Wang, P.; Wen, Y.; Kim,
  D.I.
\newblock A survey on consensus mechanisms and mining strategy management in
  blockchain networks.
\newblock {\em IEEE Access} {\bf 2019}, {\em 7},~22328--22370.

\bibitem[Benisi \em{et~al.}(2020)Benisi, Aminian, and
  Javadi]{benisi2020blockchain}
Benisi, N.Z.; Aminian, M.; Javadi, B.
\newblock Blockchain-based decentralized storage networks: A survey.
\newblock {\em Journal of Network and Computer Applications} {\bf 2020}, p.
  102656.

\bibitem[Liu \em{et~al.}(2018)Liu, Li, Chen, Hou, Xiang, and Wang]{liu2018data}
Liu, J.; Li, B.; Chen, L.; Hou, M.; Xiang, F.; Wang, P.
\newblock A Data Storage Method Based on Blockchain for Decentralization DNS.
\newblock  2018 IEEE Third International Conference on Data Science in
  Cyberspace (DSC). IEEE,  2018, pp. 189--196.

\bibitem[Di~Pierro(2017)]{di2017blockchain}
Di~Pierro, M.
\newblock What is the blockchain?
\newblock {\em Computing in Science \& Engineering} {\bf 2017}, {\em
  19},~92--95.

\bibitem[Nakamoto(2019)]{nakamoto2019bitcoin}
Nakamoto, S.
\newblock Bitcoin: A peer-to-peer electronic cash system.
\newblock Technical report, Manubot,  2019.

\bibitem[Zheng \em{et~al.}(2017)Zheng, Xie, Dai, Chen, and
  Wang]{zheng2017overview}
Zheng, Z.; Xie, S.; Dai, H.; Chen, X.; Wang, H.
\newblock An overview of blockchain technology: Architecture, consensus, and
  future trends.
\newblock  2017 IEEE international congress on big data (BigData congress).
  IEEE,  2017, pp. 557--564.

\bibitem[Dai \em{et~al.}(2019)Dai, Zheng, and Zhang]{dai2019blockchain}
Dai, H.N.; Zheng, Z.; Zhang, Y.
\newblock Blockchain for internet of things: A survey.
\newblock {\em IEEE Internet of Things Journal} {\bf 2019}, {\em
  6},~8076--8094.

\bibitem[Luo \em{et~al.}(2020)Luo, Chen, Yu, and Tang]{luo2020blockchain}
Luo, J.; Chen, Q.; Yu, F.R.; Tang, L.
\newblock Blockchain-Enabled Software-Defined Industrial Internet of Things
  with Deep Reinforcement Learning.
\newblock {\em IEEE Internet of Things Journal} {\bf 2020}.

\bibitem[Xu \em{et~al.}(2019)Xu, Su, Dai, and Yu]{xu2019apis}
Xu, Q.; Su, Z.; Dai, M.; Yu, S.
\newblock APIS: Privacy-Preserving Incentive for Sensing Task Allocation in
  Cloud and Edge-Cooperation Mobile Internet of Things with SDN.
\newblock {\em IEEE Internet of Things Journal} {\bf 2019}.

\bibitem[Wu \em{et~al.}(2019)Wu, Wang, Cai, Guo, Guo, and
  Rong]{wu2019comprehensive}
Wu, M.; Wang, K.; Cai, X.; Guo, S.; Guo, M.; Rong, C.
\newblock A comprehensive survey of blockchain: From theory to IoT applications
  and beyond.
\newblock {\em IEEE Internet of Things Journal} {\bf 2019}, {\em
  6},~8114--8154.

\bibitem[Karandikar \em{et~al.}(2019)Karandikar, Chakravorty, and
  Rong]{karandikar2019transactive}
Karandikar, N.; Chakravorty, A.; Rong, C.
\newblock Transactive Energy on Hyperledger Fabric.
\newblock  2019 Sixth International Conference on Internet of Things: Systems,
  Management and Security (IOTSMS). IEEE,  2019, pp. 539--546.

\bibitem[Xu \em{et~al.}(2017)Xu, Weber, Staples, Zhu, Bosch, Bass, Pautasso,
  and Rimba]{xu2017taxonomy}
Xu, X.; Weber, I.; Staples, M.; Zhu, L.; Bosch, J.; Bass, L.; Pautasso, C.;
  Rimba, P.
\newblock A taxonomy of blockchain-based systems for architecture design.
\newblock  2017 IEEE International Conference on Software Architecture (ICSA).
  IEEE,  2017, pp. 243--252.

\bibitem[Androulaki \em{et~al.}(2018)Androulaki, Barger, Bortnikov, Cachin,
  Christidis, De~Caro, Enyeart, Ferris, Laventman, Manevich,
  et~al.]{androulaki2018hyperledger}
Androulaki, E.; Barger, A.; Bortnikov, V.; Cachin, C.; Christidis, K.; De~Caro,
  A.; Enyeart, D.; Ferris, C.; Laventman, G.; Manevich, Y.; others.
\newblock Hyperledger fabric: a distributed operating system for permissioned
  blockchains.
\newblock  Proceedings of the Thirteenth EuroSys Conference. ACM,  2018, p.~30.

\bibitem[{Hyperledger Fabric}(2019)]{Hyperledger}
{Hyperledger Fabric}.
\newblock {Private Data},  2019.

\bibitem[Dhillon \em{et~al.}(2017)Dhillon, Metcalf, and
  Hooper]{dhillon2017blockchain}
Dhillon, V.; Metcalf, D.; Hooper, M.
\newblock {\em Blockchain Enabled Applications: Understand the Blockchain
  Ecosystem and How to Make it Work for You}; Springer,  2017.

\bibitem[Zhao \em{et~al.}(2019)Zhao, Yang, and Luo]{zhao2019consensus}
Zhao, W.; Yang, S.; Luo, X.
\newblock On Consensus in Public Blockchains.
\newblock  Proceedings of the 2019 International Conference on Blockchain
  Technology,  2019, pp. 1--5.

\bibitem[Zdrnja(2006)]{zdrnja2006security}
Zdrnja, B.
\newblock Security Monitoring of DNS traffic.
\newblock {\em University of Auckland} {\bf 2006}.

\bibitem[Govil and Govil(2007)]{govil20074g}
Govil, J.; Govil, J.
\newblock 4G mobile communication systems: Turns, trends and transition.
\newblock  2007 International Conference on Convergence Information Technology
  (ICCIT 2007). IEEE,  2007, pp. 13--18.

\bibitem[Xu \em{et~al.}(2002)Xu, Fan, Ammar, and Moon]{xu2002prefix}
Xu, J.; Fan, J.; Ammar, M.H.; Moon, S.B.
\newblock Prefix-preserving ip address anonymization: Measurement-based
  security evaluation and a new cryptography-based scheme.
\newblock  10th IEEE International Conference on Network Protocols, 2002.
  Proceedings. IEEE,  2002, pp. 280--289.

\bibitem[Kountouras \em{et~al.}(2016)Kountouras, Kintis, Lever, Chen, Nadji,
  Dagon, Antonakakis, and Joffe]{kountouras2016enabling}
Kountouras, A.; Kintis, P.; Lever, C.; Chen, Y.; Nadji, Y.; Dagon, D.;
  Antonakakis, M.; Joffe, R.
\newblock Enabling network security through active DNS datasets.
\newblock  International Symposium on Research in Attacks, Intrusions, and
  Defenses. Springer,  2016, pp. 188--208.

\bibitem[Liang \em{et~al.}(2017)Liang, Shetty, Tosh, Kamhoua, Kwiat, and
  Njilla]{liang2017provchain}
Liang, X.; Shetty, S.; Tosh, D.; Kamhoua, C.; Kwiat, K.; Njilla, L.
\newblock Provchain: A blockchain-based data provenance architecture in cloud
  environment with enhanced privacy and availability.
\newblock  Proceedings of the 17th IEEE/ACM international symposium on cluster,
  cloud and grid computing. IEEE Press,  2017, pp. 468--477.

\bibitem[Kalodner \em{et~al.}(2015)Kalodner, Carlsten, Ellenbogen, Bonneau, and
  Narayanan]{kalodner2015empirical}
Kalodner, H.A.; Carlsten, M.; Ellenbogen, P.; Bonneau, J.; Narayanan, A.
\newblock An Empirical Study of Namecoin and Lessons for Decentralized
  Namespace Design.
\newblock  WEIS. Citeseer,  2015.

\bibitem[Ali \em{et~al.}(2016)Ali, Nelson, Shea, and
  Freedman]{ali2016blockstack}
Ali, M.; Nelson, J.; Shea, R.; Freedman, M.J.
\newblock Blockstack: A global naming and storage system secured by
  blockchains.
\newblock  2016 $\{$USENIX$\}$ Annual Technical Conference
  ($\{$USENIX$\}$$\{$ATC$\}$ 16),  2016, pp. 181--194.

\bibitem[Yu \em{et~al.}(2020)Yu, Xue, Fan, and Guo]{yu2020dnstsm}
Yu, Z.; Xue, D.; Fan, J.; Guo, C.
\newblock DNSTSM: DNS Cache Resources Trusted Sharing Model Based on Consortium
  Blockchain.
\newblock {\em IEEE Access} {\bf 2020}, {\em 8},~13640--13650.

\bibitem[Liu and Albitz(2006)]{liu2006dns}
Liu, C.; Albitz, P.
\newblock {\em DNS and Bind}; " O'Reilly Media, Inc.",  2006.

\bibitem[Vayghan \em{et~al.}(2018)Vayghan, Saied, Toeroe, and
  Khendek]{vayghan2018deploying}
Vayghan, L.A.; Saied, M.A.; Toeroe, M.; Khendek, F.
\newblock Deploying microservice based applications with Kubernetes:
  experiments and lessons learned.
\newblock  2018 IEEE 11th international conference on cloud computing (CLOUD).
  IEEE,  2018, pp. 970--973.

\bibitem[Ager \em{et~al.}(2010)Ager, M{\"u}hlbauer, Smaragdakis, and
  Uhlig]{ager2010comparing}
Ager, B.; M{\"u}hlbauer, W.; Smaragdakis, G.; Uhlig, S.
\newblock Comparing DNS resolvers in the wild.
\newblock  Proceedings of the 10th ACM SIGCOMM conference on Internet
  measurement. ACM,  2010, pp. 15--21.

\bibitem[Fjellskal(2011)]{Gamelinux}
Fjellskal, E.
\newblock {Gamelinux Passive DNS},  2011.

\bibitem[{Hyperledger Fabric}(2019)]{certificatesFabric}
{Hyperledger Fabric}.
\newblock {Certificates Github},  2019.

\bibitem[Thakkar \em{et~al.}(2018)Thakkar, Nathan, and
  Viswanathan]{thakkar2018performance}
Thakkar, P.; Nathan, S.; Viswanathan, B.
\newblock Performance benchmarking and optimizing hyperledger fabric blockchain
  platform.
\newblock  2018 IEEE 26th International Symposium on Modeling, Analysis, and
  Simulation of Computer and Telecommunication Systems (MASCOTS). IEEE,  2018,
  pp. 264--276.

\bibitem[{Hyperledger Fabric}(2019)]{Hyperledgerdev}
{Hyperledger Fabric}.
\newblock {Chaincode for Developers},  2019.

\bibitem[Sompolinsky and Zohar(2015)]{sompolinsky2015secure}
Sompolinsky, Y.; Zohar, A.
\newblock Secure high-rate transaction processing in bitcoin.
\newblock  International Conference on Financial Cryptography and Data
  Security. Springer,  2015, pp. 507--527.

\bibitem[Bashir(2017)]{bashir2017mastering}
Bashir, I.
\newblock {\em Mastering blockchain}; Packt Publishing Ltd,  2017.

\bibitem[Kambourakis \em{et~al.}(2007)Kambourakis, Moschos, Geneiatakis, and
  Gritzalis]{kambourakis2007detecting}
Kambourakis, G.; Moschos, T.; Geneiatakis, D.; Gritzalis, S.
\newblock Detecting DNS amplification attacks.
\newblock  International Workshop on Critical Information Infrastructures
  Security. Springer,  2007, pp. 185--196.

\bibitem[Ranjan(2012)]{ranjan2012detecting}
Ranjan, S.
\newblock Detecting DNS fast-flux anomalies,  2012.
\newblock US Patent 8,260,914.

\bibitem[Schuba(1993)]{schuba1993addressing}
Schuba, C.
\newblock Addressing weaknesses in the domain name system protocol.
\newblock {\em Master's thesis, Purdue University, West Lafayette, IN} {\bf
  1993}.

\bibitem[Piscini \em{et~al.}(2017)Piscini, Dalton, and
  Kehoe]{Deloitteblockchain}
Piscini, E.; Dalton, D.; Kehoe, L.
\newblock {Blockchain and Cyber Security. Let\textquotesingle s Discuss},
  2017.

\bibitem[English \em{et~al.}(2018)English, Kim, and
  Nonaka]{Microsoftblockchain}
English, E.; Kim, A.D.; Nonaka, M.
\newblock {Advancing Blockchain Cybersecurity: Technical and Policy
  Considerations for the Financial Services Industry},  2018.

\bibitem[Ali \em{et~al.}(2017)Ali, Shea, Nelson, and
  Freedman]{ali2017blockstack}
Ali, M.; Shea, R.; Nelson, J.; Freedman, M.J.
\newblock Blockstack: A new decentralized internet.
\newblock {\em Whitepaper, May} {\bf 2017}.

\bibitem[Momjian(2001)]{momjian2001postgresql}
Momjian, B.
\newblock {\em PostgreSQL: introduction and concepts}; Vol. 192, Addison-Wesley
  New York,  2001.

\bibitem[Boettiger(2015)]{boettiger2015introduction}
Boettiger, C.
\newblock An introduction to Docker for reproducible research.
\newblock {\em ACM SIGOPS Operating Systems Review} {\bf 2015}, {\em
  49},~71--79.

\bibitem[{Hyperledger Fabric}(2019)]{CliHyperledger}
{Hyperledger Fabric}.
\newblock {Command-line Interface (CLI)},  2019.

\bibitem[Antonakakis \em{et~al.}(2010)Antonakakis, Perdisci, Dagon, Lee, and
  Feamster]{antonakakis2010building}
Antonakakis, M.; Perdisci, R.; Dagon, D.; Lee, W.; Feamster, N.
\newblock Building a dynamic reputation system for dns.
\newblock  USENIX security symposium,  2010, pp. 273--290.

\bibitem[Bilge \em{et~al.}(2011)Bilge, Kirda, Kruegel, and
  Balduzzi]{bilge2011exposure}
Bilge, L.; Kirda, E.; Kruegel, C.; Balduzzi, M.
\newblock EXPOSURE: Finding Malicious Domains Using Passive DNS Analysis.
\newblock  Ndss,  2011, pp. 1--17.

\end{thebibliography}

\end{document}